%Paper: hep-th/9205039
%From: cbs@yisun1.yukawa.kyoto-u.ac.jp ( C. Schwiebert 0774-20-7435)
%Date: Thu, 14 May 92 19:09:04 JST

%Here begins macro used in our institute.
\expandafter\ifx\csname phyzzx\endcsname\relax
 \message{It is better to use PHYZZX format than to
          \string\input\space PHYZZX}\else
 \wlog{PHYZZX macros are already loaded and are not
          \string\input\space again}%
 \endinput \fi
\catcode`\@=11 % This allows us to modify PLAIN macros.
\let\rel@x=\relax
\let\n@expand=\relax
\def\pr@tect{\let\n@expand=\noexpand}
\let\protect=\pr@tect
\let\gl@bal=\global
\newfam\cpfam
\newfam\mibfam
\newdimen\b@gheight             \b@gheight=12pt
\newcount\f@ntkey               \f@ntkey=0
\def\f@m{\afterassignment\samef@nt\f@ntkey=}
\def\samef@nt{\fam=\f@ntkey\the\textfont\f@ntkey\rel@x}
\def\setstr@t{\setbox\strutbox=\hbox{\vrule height 0.85\b@gheight
                                depth 0.35\b@gheight width\z@ }}

\font\seventeenrm =cmr17
\font\fourteenrm  =cmr10 scaled\magstep2
\font\twelverm    =cmr12

\font\ninerm      =cmr9

\font\sixrm       =cmr6
\font\seventeenbf =cmbx10 scaled\magstep3
\font\fourteenbf  =cmbx10 scaled\magstep2
\font\twelvebf    =cmbx10 scaled\magstep1
\font\ninebf      =cmbx9
\font\sixbf       =cmbx6
\font\seventeeni  =cmmi10 scaled\magstep3	    \skewchar\seventeeni='177
\font\fourteeni   =cmmi10 scaled\magstep2	    \skewchar\fourteeni='177
\font\twelvei     =cmmi10 scaled\magstep1	    \skewchar\twelvei='177
\font\ninei       =cmmi9                          \skewchar\ninei='177
\font\sixi        =cmmi6                           \skewchar\sixi='177
\font\seventeensy =cmsy10 scaled\magstep3    \skewchar\seventeensy='60
\font\fourteensy  =cmsy10 scaled\magstep2     \skewchar\fourteensy='60
\font\twelvesy    =cmsy10 scaled\magstep1       \skewchar\twelvesy='60
\font\ninesy      =cmsy9                          \skewchar\ninesy='60
\font\sixsy       =cmsy6                           \skewchar\sixsy='60
\font\seventeenex =cmex10 scaled\magstep3
\font\fourteenex  =cmex10 scaled\magstep2
\font\twelveex    =cmex10 scaled\magstep1
\font\seventeensl =cmsl10 scaled\magstep3
\font\fourteensl  =cmsl10 scaled\magstep2
\font\twelvesl    =cmsl10 scaled\magstep1

\font\ninesl      =cmsl9
\font\seventeenit =cmti10 scaled\magstep3
\font\fourteenit  =cmti10 scaled\magstep2
\font\twelveit    =cmti10 scaled\magstep1
\font\nineit      =cmti9
\font\seventeentt=cmtt10 scaled\magstep3
\font\fourteentt  =cmtt10 scaled\magstep2
\font\twelvett    =cmtt10 scaled\magstep1

\font\seventeencp =cmcsc10 scaled\magstep3
\font\fourteencp  =cmcsc10 scaled\magstep2
\font\twelvecp    =cmcsc10 scaled\magstep1
\font\tencp       =cmcsc10
\font\fourteencmssbx=cmssbx10 scaled\magstep2
\font\twelvecmssbx=cmssbx10 scaled\magstep1
\font\twelvecmss  =cmss12

\font\eightcmss   =cmss8
\font\seventeenmib=cmmib10 scaled\magstep3  \skewchar\seventeenmib='177
\font\fourteenmib =cmmib10 scaled\magstep2    \skewchar\fourteenmib='177
\font\twelvemib   =cmmib10 scaled\magstep1	    \skewchar\twelvemib='177
\font\elevenmib   =cmmib10 scaled\magstephalf   \skewchar\elevenmib='177
\font\tenmib      =cmmib10			    \skewchar\tenmib='177
\font\seventeenbsy=cmbsy10 scaled\magstep3   \skewchar\seventeenbsy='60
\font\fourteenbsy  =cmbsy10 scaled\magstep2     \skewchar\fourteenbsy='60
\font\twelvebsy    =cmbsy10 scaled\magstep1	      \skewchar\twelvebsy='60
\font\elevenbsy    =cmbsy10 scaled\magstephalf    \skewchar\elevenbsy='60
\font\tenbsy       =cmbsy10			      \skewchar\tenbsy='60
\begingroup
\newcount\mibchar
\mibchar=\mibfam \multiply\mibchar by 256
\advance\mibchar by 11
\global\mathchardef\bfalpha  =\mibchar \advance\mibchar by 1
\global\mathchardef\bfbeta   =\mibchar \advance\mibchar by 1
\global\mathchardef\bfgamma  =\mibchar \advance\mibchar by 1
\global\mathchardef\bfdelta  =\mibchar \advance\mibchar by 1
\global\mathchardef\bfepsilon=\mibchar \advance\mibchar by 1
\global\mathchardef\bfzeta   =\mibchar \advance\mibchar by 1
\global\mathchardef\bfeta    =\mibchar \advance\mibchar by 1
\global\mathchardef\bftheta  =\mibchar \advance\mibchar by 1
\global\mathchardef\bfiota   =\mibchar \advance\mibchar by 1
\global\mathchardef\bfkappa  =\mibchar \advance\mibchar by 1
\global\mathchardef\bflambda =\mibchar \advance\mibchar by 1
\global\mathchardef\bfmu     =\mibchar \advance\mibchar by 1
\global\mathchardef\bfnu     =\mibchar \advance\mibchar by 1
\global\mathchardef\bfxi     =\mibchar \advance\mibchar by 1
\global\mathchardef\bfpi     =\mibchar \advance\mibchar by 1
\global\mathchardef\bfrho    =\mibchar \advance\mibchar by 1
\global\mathchardef\bfsigma  =\mibchar \advance\mibchar by 1
\global\mathchardef\bftau    =\mibchar \advance\mibchar by 1
\global\mathchardef\bfupsilon=\mibchar \advance\mibchar by 1
\global\mathchardef\bfphi    =\mibchar \advance\mibchar by 1
\global\mathchardef\bfchi    =\mibchar \advance\mibchar by 1
\global\mathchardef\bfpsi    =\mibchar \advance\mibchar by 1
\global\mathchardef\bfomega  =\mibchar\endgroup
\def\seventeenf@nts{\relax
    \textfont0=\seventeenrm         \scriptfont0=\twelverm
    \scriptscriptfont0=\ninerm
    \textfont1=\seventeeni          \scriptfont1=\twelvei
    \scriptscriptfont1=\ninei
    \textfont2=\seventeensy         \scriptfont2=\twelvesy
    \scriptscriptfont2=\ninesy
    \textfont3=\seventeenex     \scriptfont3=\seventeenex
    \scriptscriptfont3=\seventeenex
    \textfont\itfam=\seventeenit
    \textfont\slfam=\seventeensl \scriptfont\slfam=\twelvesl
    \textfont\bffam=\seventeenbf
    \scriptfont\bffam=\twelvebf  \scriptscriptfont\bffam=\ninebf
    \textfont\ttfam=\seventeentt
    \textfont\cpfam=\seventeencp
    \textfont\mibfam=\seventeenmib      \scriptfont\mibfam=\twelvemib
    \scriptscriptfont\mibfam=\tenmib }
\def\fourteenf@nts{\relax
    \textfont0=\fourteenrm          \scriptfont0=\tenrm
      \scriptscriptfont0=\sevenrm
    \textfont1=\fourteeni           \scriptfont1=\teni
      \scriptscriptfont1=\seveni
    \textfont2=\fourteensy          \scriptfont2=\tensy
      \scriptscriptfont2=\sevensy
    \textfont3=\fourteenex          \scriptfont3=\twelveex
      \scriptscriptfont3=\tenex
    \textfont\itfam=\fourteenit     \scriptfont\itfam=\tenit
    \textfont\slfam=\fourteensl     \scriptfont\slfam=\tensl
    \textfont\bffam=\fourteenbf     \scriptfont\bffam=\tenbf
      \scriptscriptfont\bffam=\sevenbf
    \textfont\ttfam=\fourteentt
    \textfont\cpfam=\fourteencp
    \textfont\mibfam=\fourteenmib      \scriptfont\mibfam=\tenmib
    \scriptscriptfont\mibfam=\tenmib }
\def\twelvef@nts{\relax
    \textfont0=\twelverm          \scriptfont0=\ninerm
      \scriptscriptfont0=\sixrm
    \textfont1=\twelvei           \scriptfont1=\ninei
      \scriptscriptfont1=\sixi
    \textfont2=\twelvesy           \scriptfont2=\ninesy
      \scriptscriptfont2=\sixsy
    \textfont3=\twelveex          \scriptfont3=\tenex
      \scriptscriptfont3=\tenex
    \textfont\itfam=\twelveit     \scriptfont\itfam=\nineit
    \textfont\slfam=\twelvesl     \scriptfont\slfam=\ninesl
    \textfont\bffam=\twelvebf     \scriptfont\bffam=\ninebf
      \scriptscriptfont\bffam=\sixbf
    \textfont\ttfam=\twelvett
    \textfont\cpfam=\twelvecp
    \textfont\mibfam=\twelvemib	    \scriptfont\mibfam=\tenmib
    \scriptscriptfont\mibfam=\tenmib }
\def\tenf@nts{\relax
    \textfont0=\tenrm          \scriptfont0=\sevenrm
      \scriptscriptfont0=\fiverm
    \textfont1=\teni           \scriptfont1=\seveni
      \scriptscriptfont1=\fivei
    \textfont2=\tensy          \scriptfont2=\sevensy
      \scriptscriptfont2=\fivesy
    \textfont3=\tenex          \scriptfont3=\tenex
      \scriptscriptfont3=\tenex
    \textfont\itfam=\tenit     \scriptfont\itfam=\seveni  % no \sevenit
    \textfont\slfam=\tensl     \scriptfont\slfam=\sevenrm % no \sevensl
    \textfont\bffam=\tenbf     \scriptfont\bffam=\sevenbf
      \scriptscriptfont\bffam=\fivebf
    \textfont\ttfam=\tentt
    \textfont\cpfam=\tencp
    \textfont\mibfam=\tenmib   \scriptfont\mibfam=\tenmib
    \scriptscriptfont\mibfam=\tenmib }

\def\rm{\n@expand\f@m0 }
\def\mit{\n@expand\f@m1 }         
\def\cal{\n@expand\f@m2}
\def\it{\n@expand\f@m\itfam}
\def\sl{\n@expand\f@m\slfam}
\def\bf{\n@expand\f@m\bffam}
\def\tt{\n@expand\f@m\ttfam}
\def\caps{\n@expand\f@m\cpfam}    
\def\mib{\n@expand\f@m\mibfam}
\def\em@{\rel@x\ifnum\f@ntkey=0 \it \else
        \ifnum\f@ntkey=\bffam\it\else\rm\fi \fi }
\def\em{\n@expand\em@}
\def\seventeenpoint{\seventeenf@nts \samef@nt \b@gheight=14pt \setstr@t }
\def\fourteenpoint{\fourteenf@nts \samef@nt \b@gheight=14pt \setstr@t }
\def\twelvepoint{\twelvef@nts \samef@nt \b@gheight=12pt \setstr@t }
\def\tenpoint{\tenf@nts \samef@nt \b@gheight=10pt \setstr@t }
\normalbaselineskip = 20pt plus 0.2pt minus 0.1pt
\normallineskip = 1.5pt plus 0.1pt minus 0.1pt
\normallineskiplimit = 1.5pt
\newskip\normaldisplayskip
\normaldisplayskip = 20pt plus 5pt minus 10pt
\newskip\normaldispshortskip
\normaldispshortskip = 6pt plus 5pt
\newskip\normalparskip
\normalparskip = 6pt plus 2pt minus 1pt
\newskip\skipregister
\skipregister = 5pt plus 2pt minus 1.5pt
\newif\ifsingl@
\newif\ifdoubl@
\newif\iftwelv@  \twelv@true
\def\singlespace{\singl@true\doubl@false\spaces@t}
\def\doublespace{\singl@false\doubl@true\spaces@t}
\def\normalspace{\singl@false\doubl@false\spaces@t}
\def\Tenpoint{\tenpoint\twelv@false\spaces@t}
\def\Twelvepoint{\twelvepoint\twelv@true\spaces@t}
\def\spaces@t{\rel@x
      \iftwelv@ \ifsingl@\subspaces@t3:4;\else\subspaces@t1:1;\fi
       \else \ifsingl@\subspaces@t3:5;\else\subspaces@t4:5;\fi \fi
      \ifdoubl@ \multiply\baselineskip by 5
         \divide\baselineskip by 4 \fi }
\def\subspaces@t#1:#2;{
      \baselineskip = \normalbaselineskip
      \multiply\baselineskip by #1 \divide\baselineskip by #2
      \lineskip = \normallineskip
      \multiply\lineskip by #1 \divide\lineskip by #2
      \lineskiplimit = \normallineskiplimit
      \multiply\lineskiplimit by #1 \divide\lineskiplimit by #2
      \parskip = \normalparskip
      \multiply\parskip by #1 \divide\parskip by #2
      \abovedisplayskip = \normaldisplayskip
      \multiply\abovedisplayskip by #1 \divide\abovedisplayskip by #2
      \belowdisplayskip = \abovedisplayskip
      \abovedisplayshortskip = \normaldispshortskip
      \multiply\abovedisplayshortskip by #1
        \divide\abovedisplayshortskip by #2
      \belowdisplayshortskip = \abovedisplayshortskip
      \advance\belowdisplayshortskip by \belowdisplayskip
      \divide\belowdisplayshortskip by 2
      \smallskipamount = \skipregister
      \multiply\smallskipamount by #1 \divide\smallskipamount by #2
      \medskipamount = \smallskipamount \multiply\medskipamount by 2
      \bigskipamount = \smallskipamount \multiply\bigskipamount by 4 }
\def\normalbaselines{ \baselineskip=\normalbaselineskip
   \lineskip=\normallineskip \lineskiplimit=\normallineskip
   \iftwelv@\else \multiply\baselineskip by 4 \divide\baselineskip by 5
     \multiply\lineskiplimit by 4 \divide\lineskiplimit by 5
     \multiply\lineskip by 4 \divide\lineskip by 5 \fi }
\Twelvepoint  % That's the default
\interlinepenalty=50
\interfootnotelinepenalty=5000
\predisplaypenalty=9000
\postdisplaypenalty=500
\hfuzz=1pt
\vfuzz=0.2pt
\newdimen\HOFFSET  \HOFFSET=0pt
\newdimen\VOFFSET  \VOFFSET=0pt
\newdimen\HSWING   \HSWING=0pt
\dimen\footins=8in
\newskip\pagebottomfiller
\pagebottomfiller=\z@ plus \z@ minus \z@
\def\pagecontents{
   \ifvoid\topins\else\unvbox\topins\vskip\skip\topins\fi
   \dimen@ = \dp255 \unvbox255
   \vskip\pagebottomfiller
   \ifvoid\footins\else\vskip\skip\footins\footrule\unvbox\footins\fi
   \ifr@ggedbottom \kern-\dimen@ \vfil \fi }
\def\makeheadline{\vbox to 0pt{ \skip@=\topskip
      \advance\skip@ by -12pt \advance\skip@ by -2\normalbaselineskip
      \vskip\skip@ \line{\vbox to 12pt{}\the\headline} \vss
      }\nointerlineskip}
\def\makefootline{\baselineskip = 1.5\normalbaselineskip
                 \line{\the\footline}}
\newif\iffrontpage
\newif\ifp@genum
\def\nopagenumbers{\p@genumfalse}
\def\pagenumbers{\p@genumtrue}
\pagenumbers
\newtoks\paperheadline
\newtoks\paperfootline
\newtoks\letterheadline
\newtoks\letterfootline
\newtoks\letterinfo
\newtoks\date
\paperheadline={\hfil}
\paperfootline={\hss\iffrontpage\else\ifp@genum\tenrm
    -- \folio\ --\hss\fi\fi}
\letterheadline{\iffrontpage \hfil \else
    \rm \ifp@genum page~~\folio\fi \hfil\the\date \fi}
\letterfootline={\iffrontpage\the\letterinfo\else\hfil\fi}
\letterinfo={\hfil}
\def\monthname{\rel@x\ifcase\month 0/\or January\or February\or
   March\or April\or May\or June\or July\or August\or September\or
   October\or November\or December\else\number\month/\fi}
\def\today{\monthname~\number\day, \number\year}
\date={\today}
\headline=\paperheadline % The default is
\footline=\paperfootline % \papers
\countdef\pageno=1      \countdef\pagen@=0
\countdef\pagenumber=1  \pagenumber=1
\def\advancepageno{\gl@bal\advance\pagen@ by 1
   \ifnum\pagenumber<0 \gl@bal\advance\pagenumber by -1
    \else\gl@bal\advance\pagenumber by 1 \fi
    \gl@bal\frontpagefalse  \swing@ }
\def\folio{\ifnum\pagenumber<0 \romannumeral-\pagenumber
           \else \number\pagenumber \fi }
\def\swing@{\ifodd\pagenumber \gl@bal\advance\hoffset by -\HSWING
             \else \gl@bal\advance\hoffset by \HSWING \fi }
\def\footrule{\dimen@=\prevdepth\nointerlineskip
   \vbox to 0pt{\vskip -0.25\baselineskip \hrule width 0.35\hsize \vss}
   \prevdepth=\dimen@ }
\let\footnotespecial=\rel@x
\newdimen\footindent
\footindent=24pt
\def\Textindent#1{\noindent\llap{#1\enspace}\ignorespaces}
\def\Vfootnote#1{\insert\footins\bgroup
   \interlinepenalty=\interfootnotelinepenalty \floatingpenalty=20000
   \singl@true\doubl@false\Tenpoint
   \splittopskip=\ht\strutbox \boxmaxdepth=\dp\strutbox
   \leftskip=\footindent \rightskip=\z@skip
   \parindent=0.5\footindent \parfillskip=0pt plus 1fil
   \spaceskip=\z@skip \xspaceskip=\z@skip \footnotespecial
   \Textindent{#1}\footstrut\futurelet\next\fo@t}

\def\vfootnote#1{\Vfootnote{${#1}$}}
\def\footnote#1{\attach{#1}\vfootnote{#1}}

\let\footsymbol=\star
\newcount\lastf@@t           \lastf@@t=-1
\newcount\footsymbolcount    \footsymbolcount=0
\newif\ifPhysRev

\def\bumpfootsymbolcount{\rel@x
   \iffrontpage \bumpfootsymbolpos \else \advance\lastf@@t by 1
     \ifPhysRev \bumpfootsymbolneg \else \bumpfootsymbolpos \fi \fi
   \gl@bal\lastf@@t=\pagen@ }
\def\bumpfootsymbolpos{\ifnum\footsymbolcount <0
                            \gl@bal\footsymbolcount =0 \fi
    \ifnum\lastf@@t<\pagen@ \gl@bal\footsymbolcount=0
     \else \gl@bal\advance\footsymbolcount by 1 \fi }
\def\bumpfootsymbolneg{\ifnum\footsymbolcount >0
             \gl@bal\footsymbolcount =0 \fi
         \gl@bal\advance\footsymbolcount by -1 }
\def\fd@f#1 {\xdef\footsymbol{\mathchar"#1 }}
\def\generatefootsymbol{\ifcase\footsymbolcount \fd@f 13F \or \fd@f 279
        \or \fd@f 27A \or \fd@f 278 \or \fd@f 27B \else
        \ifnum\footsymbolcount <0 \fd@f{023 \number-\footsymbolcount }
         \else \fd@f 203 {\loop \ifnum\footsymbolcount >5
                \fd@f{203 \footsymbol } \advance\footsymbolcount by -1
                \repeat }\fi \fi }

\def\nonfrenchspacing{\sfcode`\.=3001 \sfcode`\!=3000 \sfcode`\?=3000
        \sfcode`\:=2000 \sfcode`\;=1500 \sfcode`\,=1251 }
\nonfrenchspacing
\newdimen\d@twidth
{\setbox0=\hbox{s.} \gl@bal\d@twidth=\wd0 \setbox0=\hbox{s}
        \gl@bal\advance\d@twidth by -\wd0 }
\def\removehglue{\loop \unskip \ifdim\lastskip >\z@ \repeat }
\def\roll@ver#1{\removehglue \nobreak \count255 =\spacefactor \dimen@=\z@
        \ifnum\count255 =3001 \dimen@=\d@twidth \fi
        \ifnum\count255 =1251 \dimen@=\d@twidth \fi
    \iftwelv@ \kern-\dimen@ \else \kern-0.83\dimen@ \fi
   #1\spacefactor=\count255 }
\def\step@ver#1{\rel@x \ifmmode #1\else \ifhmode
        \roll@ver{${}#1$}\else {\setbox0=\hbox{${}#1$}}\fi\fi }
\def\attach#1{\step@ver{\strut^{\mkern 2mu #1} }}
\newcount\chapternumber      \chapternumber=0
\newcount\sectionnumber      \sectionnumber=0
\newcount\equanumber         \equanumber=0
\let\chapterlabel=\rel@x
\let\sectionlabel=\rel@x
\newtoks\chapterstyle        \chapterstyle={\Number}
\newtoks\sectionstyle        \sectionstyle={\Number}
\newskip\chapterskip         \chapterskip=\bigskipamount
\newskip\sectionskip         \sectionskip=\medskipamount
\newskip\headskip            \headskip=8pt plus 3pt minus 3pt
\newdimen\chapterminspace    \chapterminspace=15pc
\newdimen\sectionminspace    \sectionminspace=10pc
\newdimen\referenceminspace  \referenceminspace=20pc
\newif\ifcn@                 \cn@true
\newif\ifcn@@                \cn@@false
\def\numberedchapters{\cn@true}
\def\unnumberedchapters{\cn@false\sequentialequations}
\def\chapterreset{\gl@bal\advance\chapternumber by 1
   \ifnum\equanumber<0 \else\gl@bal\equanumber=0\fi
   \sectionnumber=0 \let\sectionlabel=\rel@x
   \ifcn@ \gl@bal\cn@@true {\pr@tect
       \xdef\chapterlabel{\the\chapterstyle{\the\chapternumber}}}%
    \else \gl@bal\cn@@false \gdef\chapterlabel{\rel@x}\fi }
\def\@alpha#1{\count255='140 \advance\count255 by #1\char\count255}
 \def\alphabetic{\n@expand\@alpha}
\def\@Alpha#1{\count255='100 \advance\count255 by #1\char\count255}
 \def\Alphabetic{\n@expand\@Alpha}
\def\@Roman#1{\uppercase\expandafter{\romannumeral #1}}
 \def\Roman{\n@expand\@Roman}
\def\@roman#1{\romannumeral #1}    \def\roman{\n@expand\@roman}
\def\@number#1{\number #1}         \def\Number{\n@expand\@number}
\def\BLANK#1{\rel@x}               
\def\titleparagraphs{\interlinepenalty=9999
     \leftskip=0.03\hsize plus 0.22\hsize minus 0.03\hsize
     \rightskip=\leftskip \parfillskip=0pt
     \hyphenpenalty=9000 \exhyphenpenalty=9000
     \tolerance=9999 \pretolerance=9000
     \spaceskip=0.333em \xspaceskip=0.5em }
\def\titlestyle#1{\par\begingroup \titleparagraphs
     \iftwelv@\fourteenpoint\fourteenbf\else\twelvepoint\twelvebf\fi
   \noindent #1\par\endgroup }
\def\spacecheck#1{\dimen@=\pagegoal\advance\dimen@ by -\pagetotal
   \ifdim\dimen@<#1 \ifdim\dimen@>0pt \vfil\break \fi\fi}
\def\chapter#1{\par \penalty-300 \vskip\chapterskip
   \spacecheck\chapterminspace
   \chapterreset \titlestyle{\ifcn@@\chapterlabel.~\fi #1}
   \nobreak\vskip\headskip \penalty 30000
   {\pr@tect\wlog{\string\chapter\space \chapterlabel}} }

\def\section#1{\par \ifnum\lastpenalty=30000\else
   \penalty-200\vskip\sectionskip \spacecheck\sectionminspace\fi
   \gl@bal\advance\sectionnumber by 1
   {\pr@tect
   \xdef\sectionlabel{\ifcn@@ \chapterlabel.\fi
       \the\sectionstyle{\the\sectionnumber}}%
   \wlog{\string\section\space \sectionlabel}}%
   \noindent {\caps\enspace\sectionlabel.~~#1}\par
   \nobreak\vskip\headskip \penalty 30000 }
\def\subsection#1{\par
   \ifnum\the\lastpenalty=30000\else \penalty-100\smallskip \fi
   \noindent\undertext{#1}\enspace \vadjust{\penalty5000}}

\def\undertext#1{\vtop{\hbox{#1}\kern 1pt \hrule}}

\def\APPENDIX#1#2{\par\penalty-300\vskip\chapterskip
   \spacecheck\chapterminspace \chapterreset \xdef\chapterlabel{#1}
   \titlestyle{APPENDIX #2} \nobreak\vskip\headskip \penalty 30000
   \wlog{\string\Appendix~\chapterlabel} }
\def\Appendix#1{\APPENDIX{#1}{#1}}
\def\appendix{\APPENDIX{A}{}}
\def\eqname#1{\rel@x {\pr@tect
  \ifnum\equanumber<0 \xdef#1{{\rm(\number-\equanumber)}}%
     \gl@bal\advance\equanumber by -1
  \else \gl@bal\advance\equanumber by 1
   \xdef#1{{\rm(\ifcn@@ \chapterlabel.\fi \number\equanumber)}}\fi
  }#1}
\def\eq{\eqname\?}
\def\eqn{\eqno\eqname}

\def\eqinsert#1{\noalign{\dimen@=\prevdepth \nointerlineskip
   \setbox0=\hbox to\displaywidth{\hfil #1}
   \vbox to 0pt{\kern 0.5\baselineskip\hbox{$\!\box0\!$}\vss}
   \prevdepth=\dimen@}}

\def\GENITEM#1;#2{\par \hangafter=0 \hangindent=#1
    \Textindent{#2}\ignorespaces}
\outer\def\newitem#1=#2;{\gdef#1{\GENITEM #2;}}

\newdimen\itemsize                \itemsize=30pt
\newitem\item=1\itemsize;
\newitem\sitem=1.75\itemsize;     
\newitem\ssitem=2.5\itemsize;     
\outer\def\newlist#1=#2&#3&#4;{\toks0={#2}\toks1={#3}%
   \count255=\escapechar \escapechar=-1
   \alloc@0\list\countdef\insc@unt\listcount     \listcount=0
   \edef#1{\par
      \countdef\listcount=\the\allocationnumber
      \advance\listcount by 1
      \hangafter=0 \hangindent=#4
      \Textindent{\the\toks0{\listcount}\the\toks1}}
   \expandafter\expandafter\expandafter
    \edef\c@t#1{begin}{\par
      \countdef\listcount=\the\allocationnumber \listcount=1
      \hangafter=0 \hangindent=#4
      \Textindent{\the\toks0{\listcount}\the\toks1}}
   \expandafter\expandafter\expandafter
    \edef\c@t#1{con}{\par \hangafter=0 \hangindent=#4 \noindent}
   \escapechar=\count255}
\def\c@t#1#2{\csname\string#1#2\endcsname}
\newlist\point=\Number&.&1.0\itemsize;
\newlist\subpoint=(\alphabetic&)&1.75\itemsize;
\newlist\subsubpoint=(\roman&)&2.5\itemsize;

\newcount\referencecount     \referencecount=0
\newcount\lastrefsbegincount \lastrefsbegincount=0
\newif\ifreferenceopen       \newwrite\referencewrite
\newdimen\refindent          \refindent=30pt
\def\normalrefmark#1{\attach{\scriptscriptstyle [ #1 ] }}
\let\PRrefmark=\attach
\def\NPrefmark#1{\step@ver{{\;[#1]}}}
\def\refmark#1{\rel@x\ifPhysRev\PRrefmark{#1}\else\normalrefmark{#1}\fi}
\def\refend@{\refmark{\number\referencecount}}
\def\refend{\refend@{}\space }
\def\refsend{\refmark{\count255=\referencecount
   \advance\count255 by-\lastrefsbegincount
   \ifcase\count255 \number\referencecount
   \or \number\lastrefsbegincount,\number\referencecount
   \else \number\lastrefsbegincount-\number\referencecount \fi}\space }
\def\REFNUM#1{\rel@x \gl@bal\advance\referencecount by 1
    \xdef#1{\the\referencecount }}
\def\Refnum#1{\REFNUM #1\refend@ } 
\def\REF#1{\REFNUM #1\R@FWRITE\ignorespaces}
\def\Ref#1{\Refnum #1\REFWRITE }
\def\ref{\Ref\?}
\def\REFS#1{\REFNUM #1\gl@bal\lastrefsbegincount=\referencecount
    \REFWRITE }

\def\r@fitem#1{\par \hangafter=0 \hangindent=\refindent \Textindent{#1}}
\def\refitem#1{\r@fitem{#1.}}
\def\NPrefitem#1{\r@fitem{[#1]}}
\def\NPrefs{\let\refmark=\NPrefmark \let\refitem=NPrefitem}
\def\REFWRITE{\R@FWRITE\rel@x }
\def\R@FWRITE#1{\ifreferenceopen \else \gl@bal\referenceopentrue
     \immediate\openout\referencewrite=\jobname.refs
     \toks@={\begingroup \refoutspecials \catcode`\^^M=10 }%
     \immediate\write\referencewrite{\the\toks@}\fi
    \immediate\write\referencewrite{\noexpand\refitem %
                                    {\the\referencecount}}%
    \p@rse@ndwrite \referencewrite #1}
\begingroup
 \catcode`\^^M=\active \let^^M=\relax %
 \gdef\p@rse@ndwrite#1#2{\begingroup \catcode`\^^M=12 \newlinechar=`\^^M%
         \chardef\rw@write=#1\sc@nlines#2}%
 \gdef\sc@nlines#1#2{\sc@n@line \g@rbage #2^^M\endsc@n \endgroup #1}%
 \gdef\sc@n@line#1^^M{\expandafter\toks@\expandafter{\deg@rbage #1}%
         \immediate\write\rw@write{\the\toks@}%
         \futurelet\n@xt \sc@ntest }%
\endgroup
\def\sc@ntest{\ifx\n@xt\endsc@n \let\n@xt=\rel@x
       \else \let\n@xt=\sc@n@notherline \fi \n@xt }
\def\sc@n@notherline{\sc@n@line \g@rbage }
\def\deg@rbage#1{}
\let\g@rbage=\relax    \let\endsc@n=\relax
\def\refout{\par\penalty-400\vskip\chapterskip
   \spacecheck\referenceminspace
   \ifreferenceopen \Closeout\referencewrite \referenceopenfalse \fi
   \line{\fourteenrm\hfil REFERENCES\hfil}\vskip\headskip
   \input \jobname.refs
   }
\def\refoutspecials{\sfcode`\.=1000 \interlinepenalty=1000
         \rightskip=\z@ plus 1em minus \z@ }
\def\Closeout#1{\toks0={\par\endgroup}\immediate\write#1{\the\toks0}%
   \immediate\closeout#1}
\newcount\figurecount     \figurecount=0
\newcount\tablecount      \tablecount=0
\newif\iffigureopen       \newwrite\figurewrite
\newif\iftableopen        \newwrite\tablewrite
\def\FIGNUM#1{\rel@x \gl@bal\advance\figurecount by 1
    \xdef#1{\the\figurecount}}
\def\FIGURE#1{\FIGNUM #1\F@GWRITE\ignorespaces }

\def\figitem#1{\r@fitem{#1)}}
\def\FIGWRITE{\F@GWRITE\rel@x }
\def\TABNUM#1{\rel@x \gl@bal\advance\tablecount by 1
    \xdef#1{\the\tablecount}}
\def\TABLE#1{\TABNUM #1\T@BWRITE\ignorespaces }

\def\tabitem#1{\r@fitem{#1:}}
\def\TABWRITE{\T@BWRITE\rel@x }
\def\F@GWRITE#1{\iffigureopen \else \gl@bal\figureopentrue
     \immediate\openout\figurewrite=\jobname.figs
     \toks@={\begingroup \catcode`\^^M=10 }%
     \immediate\write\figurewrite{\the\toks@}\fi
    \immediate\write\figurewrite{\noexpand\figitem %
                                 {\the\figurecount}}%
    \p@rse@ndwrite \figurewrite #1}
\def\T@BWRITE#1{\iftableopen \else \gl@bal\tableopentrue
     \immediate\openout\tablewrite=\jobname.tabs
     \toks@={\begingroup \catcode`\^^M=10 }%
     \immediate\write\tablewrite{\the\toks@}\fi
    \immediate\write\tablewrite{\noexpand\tabitem %
                                 {\the\tablecount}}%
    \p@rse@ndwrite \tablewrite #1}
\def\figout{\par\penalty-400
   \vskip\chapterskip\spacecheck\referenceminspace
   \iffigureopen \Closeout\figurewrite \figureopenfalse \fi
   \line{\fourteenrm\hfil FIGURE CAPTIONS\hfil}\vskip\headskip
   \input \jobname.figs
   }
\def\tabout{\par\penalty-400
   \vskip\chapterskip\spacecheck\referenceminspace
   \iftableopen \Closeout\tablewrite \tableopenfalse \fi
   \line{\fourteenrm\hfil TABLE CAPTIONS\hfil}\vskip\headskip
   \input \jobname.tabs
   }
\newbox\picturebox
\def\p@cht{\ht\picturebox }
\def\p@cwd{\wd\picturebox }
\def\p@cdp{\dp\picturebox }
\newdimen\xshift
\newdimen\yshift
\newdimen\captionwidth
\newskip\captionskip
\captionskip=15pt plus 5pt minus 3pt
\def\fullwidth{\captionwidth=\hsize }
\newtoks\Caption
\newif\ifcaptioned
\newif\ifselfcaptioned
\def\caption{\captionedtrue \Caption }
\newcount\linesabove
\newif\iffileexists
\newtoks\picfilename
\def\fil@#1 {\fileexiststrue \picfilename={#1}}
\def\file#1{\if=#1\let\n@xt=\fil@ \else \def\n@xt{\fil@ #1}\fi \n@xt }
\def\pl@t{\begingroup \pr@tect
    \setbox\picturebox=\hbox{}\fileexistsfalse
    \let\height=\p@cht \let\width=\p@cwd \let\depth=\p@cdp
    \xshift=\z@ \yshift=\z@ \captionwidth=\z@
    \Caption={}\captionedfalse
    \linesabove =0 \picturedefault }
\def\plot{\pl@t \selfcaptionedfalse }
\def\Picture#1{\gl@bal\advance\figurecount by 1
    \xdef#1{\the\figurecount}\pl@t \selfcaptionedtrue }

\def\s@vepicture{\iffileexists \parsefilename \redopicturebox \fi
   \ifdim\captionwidth>\z@ \else \captionwidth=\p@cwd \fi
   \xdef\lastpicture{%
      \iffileexists%
         \setbox0=\hbox{\raise\the\yshift \vbox{%
              \moveright\the\xshift\hbox{\picturedefinition}}}%
      \else%
         \setbox0=\hbox{}%
      \fi
      \ht0=\the\p@cht \wd0=\the\p@cwd \dp0=\the\p@cdp
      \vbox{\hsize=\the\captionwidth%
            \line{\hss\box0 \hss }%
            \ifcaptioned%
               \vskip\the\captionskip \noexpand\Tenpoint
               \ifselfcaptioned%
                   Figure~\the\figurecount.\enspace%
               \fi%
               \the\Caption%
           \fi%
           }%
      }%
      \endgroup%
}
\let\endpicture=\s@vepicture
\def\savepicture#1{\s@vepicture \global\let#1=\lastpicture }
\def\displaypicture{\fullwidth \s@vepicture $$\lastpicture $${}}
\def\toppicture{\fullwidth \s@vepicture \topinsert
    \lastpicture \medskip \endinsert }
\def\midpicture{\fullwidth \s@vepicture \midinsert
    \lastpicture \endinsert }
\def\leftpicture{\pres@tpicture
    \dimen@i=\hsize \advance\dimen@i by -\dimen@ii
    \setbox\picturebox=\hbox to \hsize {\box0 \hss }%
    \wr@paround }
\def\rightpicture{\pres@tpicture
    \dimen@i=\z@
    \setbox\picturebox=\hbox to \hsize {\hss \box0 }%
    \wr@paround }
\def\pres@tpicture{\gl@bal\linesabove=\linesabove
    \s@vepicture \setbox\picturebox=\vbox{
         \kern \linesabove\baselineskip \kern 0.3\baselineskip
         \lastpicture \kern 0.3\baselineskip }%
    \dimen@=\p@cht \dimen@i=\dimen@
    \advance\dimen@i by \pagetotal
    \par \ifdim\dimen@i>\pagegoal \vfil\break \fi
    \dimen@ii=\hsize
    \advance\dimen@ii by -\parindent \advance\dimen@ii by -\p@cwd
    \setbox0=\vbox to\z@{\kern-\baselineskip \unvbox\picturebox \vss }}
\def\wr@paround{\Caption={}\count255=1
    \loop \ifnum \linesabove >0
         \advance\linesabove by -1 \advance\count255 by 1
         \advance\dimen@ by -\baselineskip
         \expandafter\Caption \expandafter{\the\Caption \z@ \hsize }%
      \repeat
    \loop \ifdim \dimen@ >\z@
         \advance\count255 by 1 \advance\dimen@ by -\baselineskip
         \expandafter\Caption \expandafter{%
             \the\Caption \dimen@i \dimen@ii }%
      \repeat
    \edef\n@xt{\parshape=\the\count255 \the\Caption \z@ \hsize }%
    \par\noindent \n@xt \strut \vadjust{\box\picturebox }}
\let\picturedefault=\relax
\let\parsefilename=\relax
\def\redopicturebox{\let\picturedefinition=\rel@x
   \errhelp=\disabledpictures
   \errmessage{This version of TeX cannot handle pictures.  Sorry.}}
\newhelp\disabledpictures
     {You will get a blank box in place of your picture.}
\def\FRONTPAGE{\ifvoid255\else\vfill\penalty-20000\fi
   \gl@bal\pagenumber=1     \gl@bal\chapternumber=0
   \gl@bal\equanumber=0     \gl@bal\sectionnumber=0
   \gl@bal\referencecount=0 \gl@bal\figurecount=0
   \gl@bal\tablecount=0     \gl@bal\frontpagetrue
   \gl@bal\lastf@@t=0       \gl@bal\footsymbolcount=0
   \gl@bal\cn@@false }

\def\papers{\papersize\headline=\paperheadline\footline=\paperfootline}
\def\papersize{\hsize=35.2pc \vsize=52.7pc \hoffset=0.5pc \voffset=0.8pc
   \advance\hoffset by\HOFFSET \advance\voffset by\VOFFSET
   \pagebottomfiller=0pc
   \skip\footins=\bigskipamount \normalspace }
\papers  %  This is the default
\newskip\lettertopskip       \lettertopskip=20pt plus 50pt
\newskip\letterbottomskip    \letterbottomskip=\z@ plus 100pt
\newskip\signatureskip       \signatureskip=40pt plus 3pt
\def\MEMO{\lettersize \headline=\letterheadline \footline={\hfil }%
   \let\rule=\memorule \FRONTPAGE \memohead }

\def\memodate{\afterassignment\MEMO \date }
\def\memit@m#1{\smallskip \hangafter=0 \hangindent=1in
    \Textindent{\caps #1}}
\def\subject{\memit@m{Subject:}}
\def\topic{\memit@m{Topic:}}
\def\from{\memit@m{From:}}
\def\to{\rel@x \ifmmode \rightarrow \else \memit@m{To:}\fi }
\def\memorule{\medskip\hrule height 1pt\bigskip}  % default definitions
\def\memohead{\centerline{\fourteenrm MEMORANDUM}}% see phyzzx.local
\newwrite\labelswrite
\newtoks\rw@toks
\def\letters{\lettersize
   \headline=\letterheadline \footline=\letterfootline
   \immediate\openout\labelswrite=\jobname.lab}

\let\letterhead=\rel@x
\def\addressee#1{\medskip\line{\hskip 0.75\hsize plus\z@ minus 0.25\hsize
                               \the\date \hfil }%
   \vskip \lettertopskip
   \ialign to\hsize{\strut ##\hfil\tabskip 0pt plus \hsize \crcr #1\crcr}
   \writelabel{#1}\medskip \noindent\hskip -\spaceskip \ignorespaces }
\def\rwl@begin#1\cr{\rw@toks={#1\crcr}\rel@x
   \immediate\write\labelswrite{\the\rw@toks}\futurelet\n@xt\rwl@next}
\def\rwl@next{\ifx\n@xt\rwl@end \let\n@xt=\rel@x
      \else \let\n@xt=\rwl@begin \fi \n@xt}
\let\rwl@end=\rel@x
\def\writelabel#1{\immediate\write\labelswrite{\noexpand\labelbegin}
     \rwl@begin #1\cr\rwl@end
     \immediate\write\labelswrite{\noexpand\labelend}}
\newtoks\FromAddress         \FromAddress={}
\newtoks\sendername          \sendername={}
\newbox\FromLabelBox
\newdimen\labelwidth          \labelwidth=6in
\def\makelabels{\afterassignment\Makelabels \sendername=}
\def\Makelabels{\FRONTPAGE \letterinfo={\hfil } \MakeFromBox
     \immediate\closeout\labelswrite  \input \jobname.lab\vfil\eject}
\let\labelend=\rel@x
\def\labelbegin#1\labelend{\setbox0=\vbox{\ialign{##\hfil\cr #1\crcr}}
     \MakeALabel }
\def\MakeFromBox{\gl@bal\setbox\FromLabelBox=\vbox{\Tenpoint
     \ialign{##\hfil\cr \the\sendername \the\FromAddress \crcr }}}
\def\MakeALabel{\vskip 1pt \hbox{\vrule \vbox{
        \hsize=\labelwidth \hrule\bigskip
        \leftline{\hskip 1\parindent \copy\FromLabelBox}\bigskip
        \centerline{\hfil \box0 } \bigskip \hrule
        }\vrule } \vskip 1pt plus 1fil }
\def\signed#1{\par \nobreak \bigskip \dt@pfalse \begingroup
  \everycr={\noalign{\nobreak
            \ifdt@p\vskip\signatureskip\gl@bal\dt@pfalse\fi }}%
  \tabskip=0.5\hsize plus \z@ minus 0.5\hsize
  \halign to\hsize {\strut ##\hfil\tabskip=\z@ plus 1fil minus \z@\crcr
          \noalign{\gl@bal\dt@ptrue}#1\crcr }%
  \endgroup \bigskip }

\newbox\letterb@x
\def\lettertext{\par \vskip\parskip \unvcopy\letterb@x \par }
\def\multiletter{\setbox\letterb@x=\vbox\bgroup
      \everypar{\vrule height 1\baselineskip depth 0pt width 0pt }
      \singlespace \topskip=\baselineskip }
\def\letterend{\par\egroup}
\newskip\frontpageskip
\newtoks\Pubnum   %Y%\let\pubnum=\Pubnum
\newtoks\Pubtype  \let\pubtype=\Pubtype
\newif\ifp@bblock  \p@bblocktrue
\def\PH@SR@V{\doubl@true \baselineskip=24.1pt plus 0.2pt minus 0.1pt
             \parskip= 3pt plus 2pt minus 1pt }
\def\PHYSREV{\papers\PhysRevtrue\PH@SR@V}

\def\nopubblock{\p@bblockfalse}
\def\endpage{\vfil\break}
\frontpageskip=12pt plus .5fil minus 2pt
\Pubtype={}
\def\p@bblock{\begingroup \tabskip=\hsize minus \hsize
   \baselineskip=1.5\ht\strutbox \topspace-2\baselineskip
   \halign to\hsize{\strut ##\hfil\tabskip=0pt\crcr
       \the\Pubnum\crcr\the\date\crcr\the\pubtype\crcr}\endgroup}
\def\title#1{\vskip\frontpageskip\vfill
   {\fourteenbf\titlestyle{#1}}\vskip\headskip\vfill }
\def\author#1{\vskip\frontpageskip\titlestyle{\twelvecp #1}\nobreak}

\def\address#1{\par\kern 5pt \titlestyle{\twelvepoint\sl #1}}
\def\andaddress{\par\kern 5pt \centerline{\sl and} \address}

\def\abstract#1{\vfill\vskip\frontpageskip\centerline%
               {\fourteencp Abstract}\vskip\headskip#1\endpage}

\def\ie{\hbox{\it i.e.}}       
\def\eg{\hbox{\it e.g.}}       
   
\def\\{\rel@x \ifmmode \backslash \else {\tt\char`\\}\fi }
\def\sequentialequations{\rel@x \if\equanumber<0 \else
  \gl@bal\equanumber=-\equanumber \gl@bal\advance\equanumber by -1 \fi }
\def\nextline{\unskip\nobreak\hfill\break}

\def\journal#1&#2(#3){\begingroup \let\journal=\dummyj@urnal
    \unskip, \sl #1\unskip~\bf\ignorespaces #2\rm
    (\afterassignment\j@ur \count255=#3), \endgroup\ignorespaces }
\def\j@ur{\ifnum\count255<100 \advance\count255 by 1900 \fi
          \number\count255 }
\def\dummyj@urnal{%
    \toks@={Reference foul up: nested \journal macros}%
    \errhelp={Your forgot & or ( ) after the last \journal}%
    \errmessage{\the\toks@ }}

\def\topspace{\hrule height 0pt depth 0pt \vskip}

\def\Buildrel#1\under#2{\mathrel{\mathop{#2}\limits_{#1}}}
\def\becomes#1{\mathchoice{\becomes@\scriptstyle{#1}}
   {\becomes@\scriptstyle{#1}} {\becomes@\scriptscriptstyle{#1}}
   {\becomes@\scriptscriptstyle{#1}}}
\def\becomes@#1#2{\mathrel{\setbox0=\hbox{$\m@th #1{\,#2\,}$}%
        \mathop{\hbox to \wd0 {\rightarrowfill}}\limits_{#2}}}

\let\int=\intop         
\def\lsim{\mathrel{\mathpalette\@versim<}}
\def\gsim{\mathrel{\mathpalette\@versim>}}
\def\@versim#1#2{\vcenter{\offinterlineskip
        \ialign{$\m@th#1\hfil##\hfil$\crcr#2\crcr\sim\crcr } }}
\def\big#1{{\hbox{$\left#1\vbox to 0.85\b@gheight{}\right.\n@space$}}}
\def\Big#1{{\hbox{$\left#1\vbox to 1.15\b@gheight{}\right.\n@space$}}}
\def\bigg#1{{\hbox{$\left#1\vbox to 1.45\b@gheight{}\right.\n@space$}}}
\def\Bigg#1{{\hbox{$\left#1\vbox to 1.75\b@gheight{}\right.\n@space$}}}
\def\){\mskip 2mu\nobreak }
\let\sec@nt=\sec
\def\sec{\rel@x\ifmmode\let\n@xt=\sec@nt\else\let\n@xt\section\fi\n@xt}
\def\obsolete#1{\message{Macro \string #1 is obsolete.}}
\def\firstsec#1{\obsolete\firstsec \section{#1}}
\def\firstsubsec#1{\obsolete\firstsubsec \subsection{#1}}
\def\thispage#1{\obsolete\thispage \gl@bal\pagenumber=#1\frontpagefalse}
\def\thischapter#1{\obsolete\thischapter \gl@bal\chapternumber=#1}
\def\splitout{\obsolete\splitout\rel@x}
\def\prop{\obsolete\prop \propto }
\def\nextequation#1{\obsolete\nextequation \gl@bal\equanumber=#1
   \ifnum\the\equanumber>0 \gl@bal\advance\equanumber by 1 \fi}
\def\BOXITEM{\afterassigment\B@XITEM\setbox0=}
\def\B@XITEM{\par\hangindent\wd0 \noindent\box0 }
\def\phyzzx{PHY\setbox0=\hbox{Z}\copy0 \kern-0.5\wd0 \box0 X}
        
\message{ by V.K.}
\def\schapter#1{\par \penalty-300 \vskip\chapterskip
   \spacecheck\chapterminspace
   \chapterreset \titlestyle{\ifcn@@\S\ \chapterlabel.~\fi #1}
   \nobreak\vskip\headskip \penalty 30000
   {\pr@tect\wlog{\string\chapter\space \chapterlabel}} }

\def\ssection#1{\par \ifnum\lastpenalty=30000\else
   \penalty-200\vskip\sectionskip \spacecheck\sectionminspace\fi
   \gl@bal\advance\sectionnumber by 1
   {\pr@tect
   \xdef\sectionlabel{\ifcn@@ \chapterlabel.\fi
       \the\sectionstyle{\the\sectionnumber}}%
   \wlog{\string\section\space \sectionlabel}}%
   \noindent {\S \caps\thinspace\sectionlabel.~~#1}\par
   \nobreak\vskip\headskip \penalty 30000 }
\def\eqnalign{\eqname}
\def\YITPHEAD{\null\vskip -2.0cm
 \centerline{\fourteencmssbx UJI RESEARCH CENTER}
 \centerline{\twelvecmssbx YUKAWA INSTITUTE FOR THEORETICAL PHYSICS}
 \vskip-0.0cm\centerline{\twelvecmss Kyoto University, Uji 611, Japan}
  {\eightcmss\baselineskip=0.30cm\tabskip=0pt plus400pt
  \halign to \hsize{\hskip11.3cm ## & ## & ## & ## \cr
  \ &\hskip1pt PHONE    &:&0774--20--7421\cr
  \ &\hskip1pt FAX      &:&0774--33--6226\cr}}
  \vskip 7mm}
\def\YITPletter{\lettersize \let\headline=\letterheadline
        \let\footline=\letterfootline \FRONTPAGE\YITPHEAD\addressee}
\def\YITPmark{\hbox{\fourteenmib YITP\hskip0.2cm
        \elevenmib Uji\hskip0.15cm Research\hskip0.15cm Center\hfill}}
\newif\ifYITP \YITPtrue
\def\titlepage{\FRONTPAGE\papers\ifPhysRev\PH@SR@V\fi
    \ifYITP\null\vskip-1.70cm\YITPmark\vskip0.6cm\fi
   \ifp@bblock\p@bblock \else\hrule height\z@ \rel@x \fi }
\newtoks\pubnum
\Pubnum={YITP/U-\the\pubnum}
\pubnum={9?-??}

\def\YITP{\address{Uji Research Center \break
               Yukawa Institute for Theoretical Physics\break
               Kyoto University,~Uji 611,~Japan}}

\def\globaleqnumbers{\relax\if\equanumber<0\else\global\equanumber=-1\fi}
 \def\addeqno{\ifnum\equanumber<0 \global\advance\equanumber by -1
    \else \global\advance\equanumber by 1\fi}
\mathchardef\Lag="724C

\everyjob{\input yiudef \message{Good Luck}}
\message{ modified by K-I. A.}

 \def\overbar#1{\vbox{\ialign{##\crcr
           \vrule depth 2mm
           \hrulefill\vrule depth 2mm
           \crcr\noalign{\kern-1pt\vskip0.125cm\nointerlineskip}
           $\hfil\displaystyle{#1}\hfil$\crcr}}}

\def\sqr#1#2{{\vcenter{\hrule height.#2pt
      \hbox{\vrule width.#2pt height#1pt \kern#1pt
          \vrule width.#2pt}
      \hrule height.#2pt}}}

\def\Buildrel#1\under#2{\mathrel{\mathop{#2}\limits_{#1}}}
\def\llongrarrow{\hbox to 40pt{\rightarrowfill}}

\def\journals#1&#2(#3){\unskip; \sl #1~\bf #2 \rm (19#3) }
 \def\nllap#1{\hbox to-0.35em{\hskip-\hangindent#1\hss}}

\def\rslash{\partial\kern-0.026em\raise0.17ex\llap{/}%
          \kern0.026em\relax}
\def\Dslash{D\kern-0.15em\raise0.17ex\llap{/}\kern0.15em\relax}
\mathchardef\bigtilde="0365

\def\deqalign#1{\null\,\vcenter{\openup1\jot \m@th
    \ialign{\strut\hfil$\displaystyle{##}$&$\displaystyle{##}$&$
	\displaystyle{{}##}$\hfil\crcr#1\crcr}}\,}
\newcount\eqabcno \eqabcno=97 %% \char97=``a''
\newcount\sequanumber \sequanumber=0
\newif\iffirstseq \firstseqtrue
\newif\ifequationsabc \equationsabcfalse

\def\eqnameabc#1{\relax\pr@tect
    \iffirstseq\global\firstseqfalse%
        \ifnum\equanumber<0 \global\sequanumber=\number-\equanumber
           \xdef#1{{\rm(\number-\equanumber a)}}%
           \global\advance\equanumber by -1
        \else \global\advance\equanumber by 1
           \xdef#1{{\rm(\ifcn@@ \chapterlabel.\fi \number\equanumber a)}}
        \fi
    \else\global\advance\eqabcno by 1
        \ifnum\equanumber<0
           \def#1{{\rm(\number\sequanumber \char\number\eqabcno)}}%
        \else
           \xdef#1{{\rm(\ifcn@@ \chapterlabel.\fi%
                        \number\equanumber \char\number\eqabcno)}}
        \fi
    \fi}

\def\eqsname#1{\relax\pr@tect%
        \ifnum\equanumber<0
           \def#1{{\rm(\number\sequanumber)}}
        \else
           \xdef#1{{\rm(\ifcn@@ \chapterlabel.\fi \number\equanumber)}}
        \fi}

\newtoks\publevel
\publevel={Report}   % The alternatives are Internal and Preprint

\catcode`\@=12 % at signs are no longer letters

%Here begins the text
\publevel={preprint}
\pubnum={92-07}
\date={March 1992}
\titlepage
\vskip0.5cm
\title{Constant Solutions of Reflection Equations \break
and Quantum Groups}
\author{P. P. Kulish\footnote*{ On leave of absence from St.Petersburg
Branch of Steklov Mathematical Institute.}}
\address{Yukawa Institute for Theoretical Physics, \break
Kyoto University, Kyoto 606, Japan}
\author{R. Sasaki and  C. Schwiebert\footnote{**}{Feodor Lynen and JSPS Fellow.
Supported in part by Grant in aid from Ministry of Education, Science and
Culture, No. 91081.}}
\YITP
\abstract{To the Yang-Baxter equation an additional relation can be
added. This is the reflection equation which
appears in various places, with or without  spectral parameter.
For example, in factorizable scattering on a half-line,
integrable lattice models with non-periodic boundary conditions,
non-commutative differential geometry on quantum groups, etc.
We study two forms of spectral parameter independent reflection equations,
chosen by the requirement that their solutions be comodules with respect
to the quantum group coaction leaving invariant the reflection
equations. For a variety of known solutions of the Yang-Baxter equation
we give the constant solutions of the reflection equations.
Various quadratic algebras defined by the
reflection equations are also given explicitly.}

%%%%%%%%%%%%%%%%%%%%%%%%%  References  %%%%%%%%%%%%%%%%%%%%%%%%%%%%%%%%%%%%%%%

\def \AF {Alekseev A. and Faddeev L., {\it Commun. Math. Phys.} {\bf 141}
(1991) 413.}
\def \AFST {Alekseev A., Faddeev L. and Semenov-Tian-Shansky M., \lq\lq
Hidden quantum groups inside Kac-Moody algebras."
 Preprint LOMI, E-5-91 (1991) 20p. (in [\rKulE], 148.)}
\def \AFSTV {Alekseev A., Faddeev L., Semenov-Tian-Shansky M. and Volkov A.,
\lq\lq The unraveling of the quantum group structure in the WZNW theory."
Preprint CERN-TH-5981/91 (1991) 15p.}
\def \BB {Babelon O., {\it Commun. Math. Phys.} {\bf 139} (1991) 619
and in  [\rKulE], 159.}
\def \BDF {Balog J., D\c abrowski L. and Fe\'eher L., {\it Phys. Lett.}
{\bf B257} (1991) 74.}
\def \CWSSW {Carow-Watamura U., Schlieker M., Scholl M. and  Watamura S.,
 {\it Z. Phys. C} {\bf 48} (1990) 159.}
\def \Che {Cherednik I.V., {\it Teor. Matem. Fiz.} {\bf 61} (1984) 55.}
\def \Chea {Cherednik I.V., \lq\lq Degenerate affine Hecke algebras and
Yangians." Preprint, Bonn-HE-90-04(1990) 20p.; In: Proc.RIMS-91-Project
\lq\lq Infinite analysis", Ed. T.Miwa (1992).}
\def \CFFS {Corrigan E., Fairlie D., Fletcher P. and Sasaki R.,
{\it J. Math. Phys.} {\bf 31} (1990) 776.}
\def \CG {Cremmer E. and Gervais J.-L.,
 {\it Commun. Math. Phys. \bf 144} (1992) 279; (in [\rKulE], 259.)}
\def \CGa {Cremmer E. and Gervais J.-L., {\it Commun. Math. Phys. \bf 134}
(1990) 619.}
\def \DMMZ {Demidov E., Manin Yu.I., Mukhin E. and Zhdanovich D.,
{\it Prog. Theor. Phys. Suppl.} {\bf 102} (1990) 203.}
\def \Dria {Drinfeld V.G., \lq\lq Quantum groups."
In: Proc.ICM-86, {\bf 1}, Berkeley, AMS (1987) 798.}
\def \FRT {Faddeev L., Reshetikhin N. and Takhtajan L., {\it Alg. Anal.}
 {\bf 1} (1989) 178 (in Russian, English translation: Leningrad Math. J. {\bf
1} (1990) 193).}
\def \FMai{Freidel L. and Maillet J.M., {\it Phys. Lett.} {\bf B262} (1991)
278; {\bf B263} (1991) 403.}
\def \Hie {Hietarinta J.,
\lq\lq All solutions to the constant Yang-Baxter equation in 2 dimensions."
Preprint HU-TFT-91-61 (1991) 8p.}
\def \Jim {Jimbo M., {\it Lett. Math. Phys.} {\bf 10} (1985) 63;
{\bf 11} (1986) 247; \nextline
{\it Commun. Math. Phys.}{\bf 102} (1986) 537.}
\def \Jima {Jimbo M. (Ed.),
\lq\lq Yang-Baxter equation in integrable systems."
World Scientific, Singapore 1990.}
\def \KM {Kashiwara M. and Miwa T. (Eds.), \lq\lq Special Functions."
 Proc. ICM-90 Satellite Conf., Springer, Berlin 1991}
\def \KiR {Kirillov A.N. and Reshetikhin N.Yu.,
\lq\lq Representations of the $U_q(sl(2))$, $q$-orthogonal polynomials
and invariants of links." LOMI preprint, E-9-88, 1988;
In: New developments in the theory of knots, Ed. T.Kohno, WS, 1989.}
\def \KRa {Kirillov A.N. and Reshetikhin N.Yu.,
{\it Commun. Math. Phys.} {\bf 134} (1990) 421.}
\def \KuSb {Kulish P. and Sklyanin E., {\it J. Phys.A : Math. Gen.}
{\bf 24} (1991) L435.}
\def \Kuld {Kulish.P.P., {\it Zap. Nauch. Semin. LOMI} {\bf 180} (1990) 89.}
\def \KulE {Kulish P.P. (Ed.), Proc.
\lq\lq  Euler Inter. Math. Inst. on Quantum Groups." {\it Lect. Notes Math.}
 {\bf 1510} Springer, Berlin 1992 398pp.}
\def \Kulg {Kulish P.P.,
\lq\lq Quantum groups and quantum algebras as symmetries of dynamical systems."
{}~Preprint YITP/K-959 (1991) 24p.}
\def \MNR {Mezincescu L., Nepomechie R. and Rittenberg V., {\it Phys. Lett.}
 {\bf A147} (1990) 70.}
\def \MN {Mezincescu L. and Nepomechie R.,
Preprint CERN-TH. 6152/91 14p.}
\def \MoRe {Moore G. and Reshetikhin N. Yu., {\it Nucl. Phys.} {\bf
B328} (1989) 557.}
\def \Ols {Olshanskii G.,  In  [\rKulE], 104.}
\def \OY {Okado M. and Yamane H., ~ In:  \KM\ ,  289.}
\def \PeS {Perk J.H.H. and Schultz C.L.,  ~In:
\lq\lq Non-linear integrable systems -classical and quantum",
 in Jimbo M. and Miwa T. (Eds.), WS, Singapore, 1983;
in  [\rJima ], 326.}
\def \RST {Reshetikhin N. and Semenov-Tian-Shansky M.,
{\it Lett. Math. Phys.} {\bf 19} (1990) 133.}
\def \SmWZ {Schmidke W.B., Wess J. and Zumino B., {\it Z. Phys. C} {\bf 52}
 (1991) 471.}
\def \Skla {Sklyanin E.K., {\it J. Phys. A. : Math. Gen.} {\bf 21} (1988)
2375.}
\def \Sklb {Kulish P.P. and Sklyanin E.K., To be published. }
\def \Sud {Sudbery A., {\it J. Phys.A: Math. Gen.} {\bf 23} (1990) L697.}
\def \SWZ {Schirrmacher A., Wess J. and Zumino B., {\it Z. Phys. C}
{\bf 49} (1991) 317.}
\def \WZV{ Vokos S.P., Zumino B. and Wess J.,
{\it Z. Phys. C \bf 48} (1990) 65.}
\def \YGe {Yang C.N. and Ge M.-L. (Eds.) ,
\lq\lq Braid group, knot theory and statistical mechanics"
World Scientific, Singapore 1989.}
\def \Zuma {Zumino B.,
\lq\lq Introduction to the differential geometry of quantum groups."
Preprint UCB-PTH-62/91 (1991) 17p.}

%%%%%%%%%%%%%%%%%%%   Abbreviations   %%%%%%%%%%%%%%%%%%%%%%%%%%%%%%%%%%%%

\def\sl#1{$sl_q(#1)$}
\def\w{\omega}
\def\a#1#2{a_{#1#2}}
\def\pe{\ifmmode {\cal P} \else {${\cal P}$} \fi}
\def\CD{\ifmmode {\cal D} \else {${\cal D}$} \fi}
\def\Rt{\ifmmode {\widetilde R} \else {$\widetilde R$} \fi}
\def\Rti{\ifmmode {\widetilde{R\sp{-1}}} \else {$\widetilde{R\sp{-1}}$} \fi}
\def\al{\alpha}
\def\be{\beta}
\def\ga{\gamma}
\def\de{\delta}
\def\ep{\varepsilon}
\def\AA{\ifmmode {\cal A} \else {${\cal A}$} \fi}
\def\NPrefs{\let\refmark=\NPrefmark}
\NPrefs
\def\p{{\phantom U}}

%%%%%%%%%%%%%%%%%%%%%&    Chapter 1   %%%%%%%%%%%%%%%%%%%%%%%%%%%%%%%%%%
\chapter{Introduction}
The cornerstone of the quantum inverse scattering method is the
Yang-Baxter equation. It is also  important in the theory of quantum groups
\REF\rDria{\Dria} \REF\rJim{\Jim} \REF\rFRT{\FRT}
\refmark{\rDria,\rJim,\rFRT}.
The simplest physical interpretation of the Yang-Baxter equation (YBE) is
connected with factorizable scattering on a line and with the braid
group
\REF\rJima{\Jima} \REF\rYGe{\YGe}
\refmark{\rJima,\rYGe}.
These interpretations can naturally be extended to include an additional
relation.
Such an extension will be called in this paper for simplicity
Reflection Equation (RE), for it was first obtained in the study of
factorizable scattering on a half line.
Together with the scattering matrix (called $R$-matrix in this paper),
which is a solution of the YBE, there
appears an additional object describing the reflection at
the endpoint of the half line. It will be called the $K$-matrix.
Originally, the reflection equation
$$
R(\lambda -\nu)K_1(\lambda){R\ }'(\lambda+\nu)K_2(\nu) =
K_2(\nu)R\sp{''}(\lambda+\nu)
K_1(\lambda)R\sp{'''}(\lambda -\nu)
\eqn\RE
$$
was written in the \lq\lq spectral parameter" dependent form \Ref \rChe{\Che}.
This is a matrix equation for matrices acting in the tensor product
space $V_1\otimes V_2$, %($V_1\simeq V_2$),
and $K_1 =K\otimes I$, $K_2=I\otimes K$ are standard
notations
\footnote*{The suffices of the $R$-matrices like $R = R_{12}$
indicating the base space $V_1\otimes V_2$ are suppressed in most cases.}
of the quantum inverse scattering method (QISM).
The various
$R$-matrices ${R\ }'$, $R\sp{''}$ and $R\sp{'''}$ are related
to $R$ or among themselves by certain
conjugations appropriate to the specific problem.
The significance of constant solutions to the YBE
for quantum group theory and braid group representations leads us
naturally to the analysis of  RE \RE\ without spectral parameter.
One typical form of RE reads
$$
RK_1\Rt K_2 = K_2RK_1\Rt ,  \eqn\REtw
$$
where $\Rt = \pe R\pe$, and \pe is the permutation operator
of the tensor product of the two spaces $V_1\otimes V_2$.
This form has a direct interpretation in terms of the generators of the
braid group for a solid handlebody (\eg\ for genus 1 there is only one
additional generator $\tau$)
\REF\rKulE{\KulE} \REF\rKulg{\Kulg}\refmark{\rKulE,\rKulg},
since they satisfy
$$
\sigma_1\tau\sigma_1\tau = \tau\sigma_1\tau\sigma_1 .
\eqn\Breq
$$
This is just relation \REtw\ if we identify $K_1 = \tau$,
$\pe R =\sigma_1$, and use
the fact $K_1=\pe K_2 \pe$ .

The  RE \REtw\ is closely related  to the quantum group $A(R)$\refmark{\rFRT},
which is generated by
the elements $t_{ij}$ of the quantum group matrix $T$ having relations
$$
RT_1T_2 = T_2T_1R . \eqn\RTT
$$
Any solution $K$ of \REtw\ is a two-sided comodule of the quantum group
$A(R)$, \ie\ the transformed matrix
$$
\eqalign{K_T &= T K T\sp{-1} , \cr
(K_T)_{ij} &= \sum_{m,n}t_{im}k_{mn}(T\sp{-1})_{nj} , \cr} \eqn\COMtwo
$$
is also a solution of \REtw, provided that the quantum group generators
commute with the entries of the $K$-matrix
$$
[t_{ij},k_{mn}] = 0 . \eqn\TKcom
$$
The inverse of the quantum group matrix $T$ in \COMtwo\ should be
understood as the antipode map $\ga (T)$ of the quantum group as a
Hopf algebra.

A different version of a  reflection equation without spectral parameter
is given by
\Ref\rSklb{\Sklb}
$$
RK_1R\sp{t_1}K_2 = K_2R\sp{t_1}K_1R ,  \eqn\REone
$$
where the superscript $t_1$ denotes transposition w.r.t. the first space.
This RE is different as a two-sided comodule, because it requires the
transformation property
$$
\eqalign{K_T &= T K T\sp{t} , \cr
(K_T)_{ij} &= \sum_{m,n}t_{im}k_{mn}t_{jn} , \cr}   \eqn\COMone
$$
in order that $K_T$ is a solution of \REone. In \COMone\ we denoted
the transposed quantum group matrix by $T\sp t$.

One can consider a more general  form of  RE
$$
R\sp{(1)}K_1R\sp{(2)}K_2 = K_2R\sp{(3)}K_1R\sp{(4)} ,  \eqn\REgen
$$
which is related with a generalization of the QISM to the
case of non-ultralocal commutation relations and to a lattice regularized
version of the Kac-Moody algebras
\REF\rAFSTV{\AFSTV}\REF\rAFST{\AFST}\REF\rBB{\BB}\REF\rFMai{\FMai}
\refmark{\rAFSTV-\rFMai}.

In this paper, we restrict ourselves to two types of RE's, the first
is RE \REone, which will be referred to as RE1 and the other reflection
equation is \REtw, denoted as RE2.
The reason for choosing these equations is the criterion that the $K$-matrix
should be a two-sided comodule with respect to the corresponding quantum
group coaction, which leaves the reflection equation invariant.
The comodule property is closely related to the
interpretations of reflection equations without spectral parameter.
Namely, as a first interpretation
we think of the RE as giving  a set of quadratic relations for the entries
$k_{ij}$ of the $K$-matrix which define an associative algebra \AA.
The second interpretation of the RE is that it gives an equation for a
c-number $K$-matrix, whose solutions define one dimensional
representations of the
above mentioned quadratic algebra \AA \refmark{\rSklb}.
These solutions can be used to integrate non-ultralocal commutation
relations \refmark{\rAFST,\rFMai}, also for braid group representations
\refmark{\rKulg} and some integrable models
\Ref\rCG{\CG}.
These two interpretations of the RE are in fact inseparable from each
other due to the above mentioned comodule property.
Namely, starting from a c-number solution of an RE we can construct
a new $K$-matrix $K_T$ by means of \COMtwo\ or \COMone,
which is no longer a c-number due to the quantum group matrix $T$ but it
satisfies the corresponding RE, respectively. In other words $K_T$
precisely obeys the quadratic algebra \AA\ defined by the corresponding RE.

We would like to remark that different forms of RE's (with or without
spectral parameter) appeared recently in papers on various subjects:
quantum current algebras and conformal field theory
\Ref\rMoRe{\RST\nextline\MoRe},
modified Knizhnik-Zamolodchikov equations \Ref\rChea{\Chea},
integrable spin chains with non-periodic boundary conditions
\REF\rSkla{\Skla} \REF\rMNR{\MNR} \REF\rMN{\MN} \REF\rKuSb{\KuSb}
\refmark{\rSkla-\rKuSb}, non-commutative differential geometry
on quantum groups \REF\rZuma{\Zuma}\REF\rAF{\AF}\refmark{\rZuma,\rAF},
twisted Yangians \Ref\rOls{\Ols}, and so on.

This paper is organized as follows. Some of the known results for the
reflection
equations without spectral parameter are briefly reviewed in section 2.
Two types of reflection equations are selected  based on the requirement of
the two-sided comodule property under the coaction of the corresponding quantum
groups.
Section 2 also contains discussions and some new results on
various general properties of the reflection equations, their constant
solutions and the corresponding quadratic algebras  for
the $R$-matrix (spin $1/2$ representation and the universal $R$-matrix) of the
simplest quantum algebra \sl2.
Section 3 gives the new results on the constant solutions to the two types of
reflection
equations for a wide class of known constant solutions of the YBE.
The final section is devoted for summary and discussion.

The enlightening discussions with Prof. E. K. Sklyanin and
communication of his unpublished results are gratefully acknowledged.

%%%%%%%%%%%%%%%%%%%%%%%%%  Chapter 2   %%%%%%%%%%%%%%%%%%%%%%%%%%%%

\chapter{Properties of Reflection Equations}

\section{Reflection equation 1}
It was pointed out in \refmark{\rSkla}
that the quadratic algebra defined by an
appropriate reflection equation can be considered as a two-sided
comodule
$$
K_T = T K T\sp {\sigma}  \eqn\coact
$$
\noindent for the corresponding quantum group and an anti-automorphism
$\sigma$. It should be reminded  that the entries of $K$ commute with
those of $T$ as mentioned in \TKcom.
As the first example we specify $\sigma$ to be the transposition
$ \sigma:\ T \rightarrow T\sp t $, or $ \sigma(t_{ij})=t_{ji} $.
In this case the reflection equation has the form RE1 as introduced in the
previous section \REone\
$$
\hskip-4cm{\rm RE1 :} \hskip4cm R K_1 R\sp {t_1} K_2 = K_2 R\sp {t_1} K_1 R  ,
\eqn\reeqon
$$
\noindent where $t_1$ means transposition in the first space.
The invariance of this equation w.r.t. the coaction \coact\ follows from
the defining relations of the quantum group generators
$$
R T_1 T_2 = T_2 T_1 R , \eqn\rtt
$$
\noindent which can be transformed also into the forms
$$
\eqalign{R T_1\sp t T_2\sp t &= T_2\sp t T_1\sp t R , \cr
         T_2 R\sp {t_1} T_1\sp t &= T_1\sp t R\sp {t_1} T_2 ,   \cr
         T_1 R\sp {t_1} T_2\sp t &= T_2\sp t R\sp {t_1} T_1 ,  \cr}\eqn\rttt
$$
\noindent provided the $R$-matrix has the following properties ($R\sp t =
R\sp{t_1\circ t_2}$\/)
$$
R = \pe R\sp t \pe , \quad  R\sp {t_1} = \pe R\sp {t_1} \pe . \eqn\peR
$$
\noindent  They are certainly true for the \sl{N} $R$-matrix which
is the main subject of
our research, but not for its multiparameter generalizations or in some other
cases.
In case $R$-matrices do not have the properties \peR, the two-sided comodule
property \COMone\ or \coact\ is valid for
a generalized equation
$$
R K_1 R\sp {t_1} K_2 = K_2 {\widetilde {R\sp {t_1}}} K_1 {\widetilde {R\sp t}}
,
\eqn\rethree
$$
instead of the original equation \reeqon\ . In such situations we call
 \rethree\ the RE1 belonging to the given $R$-matrix.
The $R$-matrices on the right hand side are defined as
${\widetilde {R\sp {t_1}}} = \pe R\sp{t_1}\pe$ and
${\widetilde {R\sp t}}= \pe R\sp t \pe$.
For each given $R$-matrix the RE1 defines an associative quadratic algebra
which will be denoted by $\AA_1$, and similarly by $\AA_2$ for the RE2 to be
discussed shortly.

In the remainder of this section we will study in some
detail the basic example of \sl2 for both reflection equations
RE1 \REone\ , \reeqon\ and RE2 \REtw. The $R$-matrix is well known
$$
R=\pmatrix{q& & & \cr  &1& & \cr  &\w&1& \cr  & & &q\cr} , \quad
R\sp {t_1}=\pmatrix{q& & &\w\cr  &1& & \cr  & &1& \cr  & & &q\cr} ,
\quad \w=q-q\sp {-1} . \eqn\Rmat
$$
We adopt the convention
that all matrix elements not written explicitly are zeros.
The notation for the $R$-matrices is
such that the upper index pair $(ij)$ of
${R\sp {ij}}_{kl}$ numbers the rows in the natural order and likewise the lower
index pair $(kl)$ numbers the columns.
In the basis \Rmat\ the permutation operator is given by
$$ \
\pe=\pmatrix{1& & & \cr  & &1& \cr   &1& & \cr  & & &1\cr} .
\eqn\eq
$$
\noindent For the  $R$-matrix \Rmat\ the quadratic
algebra  $\AA_1 $ is generated by the
four elements of the $K$-matrix
$$
K=\pmatrix{\al&\be\cr  \ga&\de\cr} \eqn\kreon
$$
\noindent which satisfy the relations \refmark{\rSklb}
$$
\eqalign{[\al,\be]&=\w\al\ga , \cr  \al\ga&=q\sp 2 \ga\al , \cr}  \qquad
\eqalign{[\al,\de]&=\w(q\be\ga+\ga\sp 2) , \cr  [\be,\ga]&=0 , \cr}  \qquad
\eqalign{[\be,\de]&=\w\ga\de , \cr  \ga\de&=q\sp 2 \de\ga .
\cr} \eqn\Ala
$$
\noindent This algebra $\AA_1 $  has two central elements
$$ c_1 = \be - q\ga , \qquad c_2 = \al\de -q\sp2\be\ga . \eqn\qoncen
$$
\noindent Under the coaction \coact\ (or \COMone) these central elements
transform homogeneously with respect to $GL_q(2)$
$$
c_1(K_T) = (det_qT)c_1(K), \qquad c_2(K_T) = (det_qT)\sp2c_2(K) . \eqn\cetT
$$
Here $T$ is the matrix of the $GL_q(2)$ quantum group generators
satisfying the well known relations \refmark{\rJim,\rFRT} which follow from
\rtt\
$$
T=\pmatrix{a&b \cr  c&d \cr} : \qquad
\eqalign{a b&=q b a , \cr  a c &=q  c a , \cr}  \qquad
\eqalign{ b d &=q d  b , \cr   c  d &=q  d  c  , \cr}  \qquad
\eqalign{a d -q b c &= d  a-q\sp{-1} b c  , \cr   b c &= c  b . \cr}
\eqn\Qgrop
$$
\noindent For the $SL_q(2)$ case \ie\
$det_qT=a d -q b c = d  a-q\sp{-1} b c=1$, the central elements  are
invariant.
If we restrict the above quadratic algebra $\AA_1$ by requiring that
$c_1(K) = 0$ (\ie $\be =q\ga$) then \Ala\ reduces to the well known
$GL_{q\sp2}(2)$ relations, namely the quatum group $GL_q(2)$ with the parameter
$q$ replaced by $q\sp 2$
$$
\eqalign{\al\be&=q\sp2\be\al , \cr  \al\ga&=q\sp 2 \ga\al , \cr}  \qquad
\eqalign{\be\de&=q\sp2\de\be , \cr  \ga\de&=q\sp 2 \de\ga , \cr}  \qquad
\eqalign{\al\de-q\sp2\be\ga&=\de\al-q\sp{-2}\be\ga , \cr  \be\ga&=\ga\be . \cr}
\eqn\Alaqsq
$$
Therefore the second condition of the \lq\lq degeneracy\rq\rq, $c_2=0$,
simply corresponds to the  vanishing quantum determinant of the above
mentioned \lq\lq$GL_{q\sp2}(2)$\rq\rq.
It is interesting to note that the above relations with $\be =q\ga$ are also
satisfied  by $K = TT\sp t $ , which is obtained by \COMone\ from a c-number
solution $K=1$ (the  $2\times2$ unit matrix) to be discussed shortly.
A similar phenomenon has been known for some time \REF\rCFFS{\CFFS}
\REF\rWZV{\WZV} \refmark{\rCFFS,\rWZV}.
Namely, if $T$ is a $GL_q(2)$ matrix then $T\sp n$ is a $GL_{q\sp n}(2)$
matrix.

Now we are interested in c-number solutions of the reflection
equation. From the algebra \Ala\ it is easy to conclude that there
are only two types of  non-trivial solutions \refmark\rSklb\ :
(i) the $K$-matrix proportional to the quantum invariant metric $\ep_q$ of
\sl2;
(ii) an upper triangular $K$-matrix
with three arbitrary parameters (excluding the irrelevant overall factor
leaves essentially two arbitrary parameters)
$$
 K\sp {(1)}\sim\ep_q=\pmatrix{ &1\cr  -q& \cr} , \quad
K\sp {(2)}=\pmatrix{\al&\be\cr  &\de\cr} . \eqn\sol
$$
It is easy to see that the $\AA_1$ relations become trivial for $\ga=0$, which
gives the c-number solution $K\sp{(2)}$. Similarly we get $K\sp{(1)}$ by
putting $\al=\de=0$.
The two-sided coaction \coact\  has $K\sp {(1)}$ as a
fixed point due to the well known fact
\refmark{\rCFFS} that $ T\ep_qT\sp t=(det_qT)\ep_q $
and that for $SL_q(2)$ the quantum determinant is unity $det_qT=1$.

It is well known that the $R$-matrix \Rmat\ is the spin $({{1}\over{2}},
{{1}\over{2}})$
representation of the universal $R$-matrix of the quasi-triangular
Hopf algebra \sl2 generated by three elements $J, X_+, X_-$
satisfying the relations
$$
\eqalignno{[J,X_{\pm}] &= \pm X_{\pm} , &\eqnalign{\Xpm} \cr
[X_+,X_-] &= [2J]_q \equiv (q\sp {2J}-q\sp {-2J})/(q-q\sp {-1}) .
&\eqnalign{\XpmJ} \cr}
$$
\noindent The universal $R$-matrix is
\REF\rKiR{\KiR\nextline\KRa}\refmark{\rDria,\rKiR}
$$
R_U = q\sp {2J \otimes J} \sum\sp {\infty}_{n=0} {{(1-q\sp {-2})\sp n}\over
{[n;q\sp {-2}]!}} f\sp n \otimes e\sp n  ,  \eqn\RU
$$
where $ [n;q]=(1-q\sp n)/(1-q) $ and
$$
e = q\sp J X_+ , \quad  f = q\sp {-J} X_- .  \eqn\eq
$$
Hence it is possible to generalize RE in the tensor product $ V_1 \otimes V_2 $
of two irreducible representations  of $q$-spin $s_1$ and
$s_2$. In the analysis of the quantum Liouville equation on a strip
\refmark{\rCG} a quasi-group like element was constructed
$$
K_U  = q\sp {J\sp 2} \sum_{n=0}\sp {\infty} {f_n \over [n]!}\ q\sp {-nJ}
(X_+)\sp n ,
\eqn\ku
$$
which just turned out to be a universal solution of RE1 for \sl2 with
universal $R$-matrix \RU\ and
$$
R_U\sp {t_1} = \sum\sp {\infty}_{n=0} {{(1-q\sp {-2})\sp n}\over{[n;q\sp 2]!}}
e\sp n \otimes e\sp n q\sp {2J \otimes J} .  \eqn\RUt
$$
In this case RE1 is understood to be defined in the tensor product of two
copies of the quantum algebras $sl_q(2)\otimes sl_q(2)$.
It is very remarkable that the universal $K$-matrix in fact depends on
one parameter only, as the coefficients $f_n$ are defined by the recurrence
relation related to that of the $q$-Hermite polynomials
$$
f_{n+1} = f_1 f_n + (q\sp {2n} - 1)f_{n-1} , \qquad f_0=1 ,  \eqn\eq
$$
and the free parameter is $f_1$. The proof that $K_U$ indeed is a
solution of RE1 follows from its property \refmark{\rCG}
$$
\Delta (K_U\sp \p) = (K_U\sp \p)_1\sp \p R_U\sp {t_1} (K_U\sp \p)_2\sp \p
\eqn\deltaK
$$
and the definition of the canonical element, i.e. the universal
$R$-matrix, for the quasi-triangular Hopf algebra. The universal
$R$-matrix intertwines the coproduct and
the permuted coproduct \refmark{\rDria}
$$
R_U\sp \p \Delta (\cdotp) = \Delta' (\cdotp) R_U\sp \p  ,  \eqn\Copro
$$
where $\Delta'$ is the permuted coproduct
$\Delta'(\cdotp)=\pe\Delta(\cdotp)\pe$.
The relations \deltaK\ and  \Copro\ lead to
$$
R_U\sp \p (K_U\sp \p)_1\sp \p R_U\sp {t_1} (K_U\sp \p)_2\sp \p = (K_U\sp
\p)_2\sp \p R_U\sp {t_1} (K_U\sp \p)_1\sp \p R_U\sp \p  \eqn\eq
$$
when taking into account that \RUt\ is invariant w.r.t. the
similarity transformation by the permutation operator (namely the second
equation of \peR\ holds for $R_U$).
The constant  solution $K\sp {(2)}$ in \sol\ corresponds to the case of the
$q$-spin $({1 \over 2},{1 \over 2})$ representation of the universal solution
\ku\ for \sl2.

Let us point out the invariance of RE1 w.r.t the transformation
$$
K' = e\sp {\kappa J} K e\sp {\kappa J} ,\qquad \kappa : {\rm arbitrary \
constant}
\eqn\JKJ
$$
which explains the difference in the number of independent parameters
of $K\sp {(2)}$ in \sol\ (two parameters) and in \ku\ (one
parameter, $f_1\,$), as it can be used to transform away one parameter
in \sol\ by properly adjusting the arbitrary constant $\kappa$.
This invariance follows from the relations
$$
[R,J_1 + J_2] = 0 , \qquad  [R\sp {t_1},J_1 - J_2] = 0 , \eqn\RJcom
$$
which are valid for the universal $R$-matrix as well. The free parameter $f_1$
of the universal solution \ku\ is closely related to the central element $c_1$
of the quadratic algebra $\AA_1$ , namely $c_1 \propto f_1$ .

The first solution $K\sp {(1)}$ in \sol\ of RE1  can also be extended
to higher representations. It is proportional to the element $w$ of
the quantum Weyl group
\refmark\rKiR,
which can be defined in each finite
dimensional representation $V_j$, dim$\,V_j=2j+1$, as
$$
w \simeq q\sp {-J} \ep_j , \qquad (\ep_j)_{m,m'} = (-1)\sp {j-m} \de_{m,-m'}
, \quad -j\leq m,m' \leq j, \eqn\eq
$$
or by its commutation properties with the generators of \sl2
$$
w J = - J w , \quad w X_{\pm} = - q\sp {\pm 1} X_{\mp} w  .  \eqn\eq
$$
It can be shown that $K_U\sp \p=w$ indeed is a solution of RE1 due to the
following properties of the universal $R$-matrix
$$
w_1\sp \p R\sp {t_1}_U w_1\sp {-1} = w_1\sp {-1} R\sp {t_1}_U w_1\sp \p = R\sp
{-1}_U  .  \eqn\eq
$$
\section{Reflection equation 2}
In the lattice version of the current algebras or the Kac-Moody algebras
\refmark{\rAFSTV,\rAFST} another type of reflection equation appears
which we denote as RE2
$$
\hskip-4cm {\rm RE2 :} \hskip4cm R K_1 \Rt K_2 = K_2 R K_1 \Rt .  \eqn\reeqtw
$$
It defines the set of relations
of an associative quadratic algebra $\AA_2$
generated by the entries of the $K$-matrix, and it is a two-sided comodule
w.r.t. the quantum group coaction \COMtwo, namely the antiautomorphism
$\sigma$ is the inverse $\sigma(T)=T\sp{-1}$. Let us again consider the
\sl2 $R$-matrix \Rmat\  and this time denote the entries of the $K$-matrix by
$$
K=\pmatrix{u&x\cr  y&z\cr} ,  \eqn\kretw
$$
then we find that the component form of the algebra \reeqtw\ reads
$$
\eqalign{ux&=q\sp {-2}xu , \cr  uy&=q\sp 2yu , \cr}  \qquad
\eqalign{[u,z]&=0 , \cr  [x,y]&=q\sp {-1}\w (uz-u\sp 2) , \cr}  \qquad
\eqalign{[x,z]&=-q\sp {-1}\w ux , \cr  [y,z]&=q\sp {-1}\w yu  .
\cr} \eqn\Alb  $$
The central elements of the above algebra $\AA_2$ are also known
\refmark\rZuma\
and they are invariant under the $GL_q(2)$ coaction in contrast to the
$SL_q(2)$ invariance for $\AA_1$,
$$
\eqalign{c_1 &= u + q\sp2 z , \cr  c_1(K_T) &= c_1(K) , \cr}  \qquad
\eqalign{c_2 &= uz -q\sp2 yx , \cr  c_2(K_T) &= c_2(K) .  \cr} \eqn\centwo
$$
The quadratic algebras \Ala\  and \Alb\  are equivalent to each other
by a simple substitution:
\ie\ the following substitution in  \Ala\
$$
\{ \al,\be,\ga,\de\} \to \{y,-z/q,u,-x/q\} \eqn\Asub
$$
produces \Alb\ in which $q$ is replaced by $1/q$.
Generally, however, this is not the case for the
quadratic algebras defined by \sl{N} $R$-matrices, as for $N>2$ there is
no relation between $T\sp t$ and the inverse (antipode) $T\sp{-1}=\ga (T)$
contrary to the \sl2 case
where we have
$$
T\sp{t} = \ep_q T\sp {-1} \ep_q\sp {-1} ,  \eqn\eq
$$
and $\ep_q$ is the $2 \times 2$ quantum metric in \sol.

The c-number solution of RE2 for the $sl_q(2)$ $R$-matrix \Rmat\ can be easily
obtained as a one-dimensional representation of the above $\AA_2$.
Similarly to the case of $\AA_1$ there are two of them,
$$
K\sp {(1)}\simeq\pmatrix{1& \cr  &1\cr} , \quad
K\sp {(2)}=\pmatrix{ & x \cr y & z \cr} . \eqn\solt
$$
A glance at the $\AA_2$ relations shows that they become trivial for $u=0$,
which gives $K\sp{(2)}$.
The above solutions  are related to the solutions \sol\ of RE1
and they are mapped to each other by the substitution mentioned above.
As in the case of RE1,  $K\sp{(1)}$ is a fixed point under the coaction of the
quantum group \COMtwo.
Obviously RE2 has always a solution proportional to the unit matrix for any
choice of the $R$-matrix. It is interesting to note that RE2 can be rewritten
into the following equivalent form \refmark{\rAFSTV}
$$
R K_1 R\sp{-1} K_2 = K_2 \Rt\sp{-1} K_1 \Rt , \qquad
\Rt\sp{-1} = \pe R\sp{-1}\pe.
\eqn\reeqtwp
$$

Next let us discuss certain invariance properties of the
\sl{N} $R$-matrices, which are important for the understanding of their
multiparametric generalizations.
For the $R$-matrices of \sl{N} the reflection equation \reeqtw\
is invariant as is
\reeqon\  w.r.t. the similarity transformation
$$
K' = e\sp {\tau J} K e\sp {-\tau J} ,  \eqn\JKmJ
$$
due to
$$
[R,J_1+J_2]=0 , \qquad  [\Rt,J_1+J_2]=0 ,  \eqn\eq
$$
where $J$ belongs to the Cartan subalgebra and $\tau$ is some parameter.
There is an additional transformation for the \sl{N} $R$-matrix
\Ref\rKuld{\Kuld}
$$
R' = U_1\sp 2 R U_2\sp {-2} ,  \eqn\traa
$$
which transforms a solution of the YBE to the solution $R'$ if the
$R$-matrix is symmetric, i.e.
$$ [R,U_1U_2]=0 , \qquad U \in {\rm Mat}({\bf C}\sp n) .  \eqn\RUUcom
$$
The corresponding transformation of the $K$-matrix is
$$
K' = U K U\sp {-1} \eqn\trab
$$
and the RE2 \reeqtw\  is invariant under \traa\  and \trab. This
transformation gives relations among solutions and quadratic algebras
of the reflection equations \reeqtw\  for the one-parameter and some
multiparameter  $R$-matrices which we will discuss in the following section.

\section{Comodule Properties}
Before concluding this section we find it instructive to remark on the
two-sided comodule properties \coact\ (or \COMtwo, \COMone) of the reflection
equations in general.
For concreteness let us choose RE1 for a generic $R$-matrix
but the situation for RE2 is essentially the same.

The two-sided comodule property simply states that given a solution $K$ of RE1
we can associate with it another solution $K_T$
(possibly in a different space) by (1.5) provided~:
(i) $T$ satisfies the FRT-relations \rtt\ and
(ii) each element of $K$ should commute with each element $t_{ij}$ of $T$.
The general solutions of the FRT-relations describe the quantum group $A(R)$.
We may say that  $K$ is a comodule with respect to the quantum group $A(R)$
with the coaction $\de$ :
$$
\de\,:\  \AA \to A(R) \otimes \AA . \eqn\gecoact
$$
The quantum group $A(R)$ is known to have various representations.
Suppose we choose one of them, $\rho(t_{ij})$ which acts on a linear space
$V_{\rho}$,
$\rho(t_{ij}) \in {\rm Mat}(V_{\rho})$, then $K_T$ is a new solution of RE1 in
a new
space
$$
{\rm Mat}(V_{\rho})\otimes K  .
$$
Of course we can treat $T$ as a general quantum group matrix without
specifying any representation as we have done  in this section. Then the
new solution $K_T$ lies in a space which is written symbolically as in
\gecoact.

Some concrete examples are in order. Suppose we have a solution of the
YBE without spectral parameter acting in $V_1\otimes V_2\otimes V_3$
($V_1 \simeq V_2$ , $V_3$ :  arbitrary)
$$
R_{12}R_{13}R_{23} = R_{23}R_{13}R_{12} .
$$
This provides a solution of the FRT-relations in which $V_3$ is considered
as the \lq\lq quantum space\rq\rq , \ie\
$$
\rho(T_1) =R_{13} . \eqn\TRonethr
$$
The  $(l , m)$ element of the representation matrix is given by
$\rho(t_{ij})\sp l{}_m = (R_{13})\sp{il}{}_{jm}$.
Thus we find a new solution $K_T$ \COMone\ of RE1, whose components are
matrices on $V_3$.

In this way we can interpret the symmetry properties of $K$ \JKJ\ which
explained the equivalence of the two solutions
\sol\ (two parameters) and \ku\ (one parameter $f_1$).
We can choose the trivial representation of $T$ in $GL_q(2)$, \Qgrop ,
$\rho_{\rm trivial}(T)={\rm diag}(a,d),~a,d \in {\bf C}$.
Then the two-sided comodule
property simply means that if $K$ is a c-number solution of the RE1 then
$$
\pmatrix{a &\cr  &d \cr}K \pmatrix{a &\cr  &d \cr} \eqn\simplCom
$$
is another c-number solution. This is the essence of the symmetry argument
\JKJ, \RJcom.

Having clarified the meaning of the two-sided comodule properties \coact\ of
RE1 and RE2 we are ready to discuss a more general form of the reflection
equations (without spectral parameter) \REgen
$$
R\sp{(1)}K_1R\sp{(2)}K_2 = K_2R\sp{(3)}K_1R\sp{(4)} .  \eqn\REgen
$$
This equation defines the relations among the generators of
an associative algebra. The algebra has the two-sided comodule property
$$
K \rightarrow K_{T,S} = T K S ,          \eqn\coagen
$$
provided that $T$ and $S$ satisfy the following two sets of equations
$$
         R\sp{(1)} T_1 T_2 = T_2 T_1 R\sp{(1)} , \qquad
         R\sp{(4)} S_1 S_2 = S_2 S_1 R\sp{(4)} ,   \eqn\trtt
$$
\vskip-1.5cm
$$
         S_1 R\sp {(2)} T_2 = T_2 R\sp {(2)} S_1 ,  \qquad
         T_1 R\sp {(3)} S_2 = S_2 R\sp {(3)} T_1 . \eqn\TRV
$$
The first set \trtt\ simply means that $T$ and $S$ belong to (possibly
different)
quantum groups specified by $R\sp{(1)}$ and $R\sp{(4)}$, respectively.
The two equations in the second set \TRV\ are compatible to each other \eg\
when $R\sp{(3)}= \Rt\sp{(2)}$. They require certain relations among
$T$ and $S$.  Namely  $R\sp{(2)}$ intertwines $S_1$ and $T_2$ and $R\sp{(3)}$
should intertwine $T_1$ and $S_2$ which are necessary for the two-sided
comodule property.

As before the elements of $T$ and $S$ should commute with the elements of $K$,
$$
[t_{ij},k_{l m}] = [s_{ij}, k_{l m}] = 0 . \eqn\TKVcom
$$
It is conceptually straightforward to give concrete examples of the two-sided
comodule transformations for the general form of RE, which  corresponds  to
\TRonethr\ (cf. \refmark{\rBB,\rFMai}).

%%%%%%%%%%%%%%%%%%%%%%%    Chapter 3    %%%%%%%%%%%%%%%%%%%%%%%%%%%%%%%

\chapter{Constant Solutions of Reflection Equations}

\section{$sl_q(2)$ spin 1 representation}
In the preceding section we have  discussed the constant
solutions to RE1 and RE2 for \sl2\ $R$-matrix.
The quantum metric type and the upper-triangular solutions were found,
and they already exhausted the list
of solutions of RE1 for \sl2.
As we have seen, both of them can be generalized to
universal solutions for RE1, therefore we expect this
type of solutions in the higher spin
representations of \sl2.  But, in principle,
there can be other solutions in these cases in addition, which are not
reductions of universal ones but tied to particular representations.

To examine this point we now discuss the constant
solutions for the spin one case.  The $R$-matrix can be derived from the
universal $R$-matrix \RU\
$$
R=\pmatrix{q\sp 2& & & & & & & & \cr   &1& & & & & & & \cr
& &q\sp {-2}& & & & & & \cr   &\Delta& &1& & & & & \cr
& &q\sp {-1}\Delta& &1& & & & \cr    & & & & &1& & & \cr
& &q\sp {-1}\w\Delta& &q\sp {-1}\Delta& &q\sp {-2}& & \cr   & & & & &\Delta&
&1& \cr
& & & & & & & &q\sp 2\cr} , \quad \Delta=(q+q\sp {-1})\w . \eqn\Rspinone
$$
By inserting this $R$-matrix into RE1 we find four constant
solutions.  Two of them are given by
$$
K\sp {(1)}=\pmatrix{& &-q\sp {-1}\cr  &1& \cr  -q& & \cr} , \quad
K\sp {(2)}=\pmatrix{\a11&\a12&h_{13}\cr  &\a22&h_{23}\cr
& &h_{33}\cr} , \eqn\spon
$$
\noindent where $h_{13},\ h_{23},\ $ and $h_{33}$ are functions of the free
parameters $\a11,\ \a12,\ $ and $\a22$ given by
$ h_{13}=\w \a22 + {\a12\sp 2}/{(q+q\sp {-1})\a11},\  h_{23}={q \a12
\a22}/{\a11},\  h_{33}={q\sp 2 \a22\sp 2}/{\a11} $.
It is clear that $ K\sp {(1)} $ corresponds
to the metric type universal solution and $ K\sp {(2)} $ to the
universal triangular solution \refmark{\rCG}.
The other two solutions have two- and one- parameter dependence  with vanishing
determinants
$$
K\sp{(3)} =\pmatrix{& & a_{23}\sp2/a_{33}Y \cr & & a_{23} \cr & & a_{33}},
\quad Y=q+1/q, \quad
K\sp{(4)} =\pmatrix{& & a_{13} \cr & & \cr & & } .
$$
At first sight they appear to be independent of the universal solution
$K\sp{(2)}$.
But one can show that $K\sp{(3)}$ can be obtained from $K\sp{(2)}$ in an
appropriate scaling limit and likewise $K\sp{(4)}$ from $K\sp{(3)}$.
So in this particular case the two types of the universal solutions contain
all the solutions and there are no additional solutions  mentioned above.
In the following we use  non-vanishing
determinants as a criterion for selecting `interesting' solutions.
We also omit the solutions which can be obtained
from the others by specializing the parameters and/or by scaling limits.

\section{$sl_q(N)$ for $N\ge3$}
Next we proceed to \sl3\ and later to \sl{N}\ for generic $N$
in the fundamental  representation.
The $R$-matrix of \sl3 is given by
$R={\rm diag}(q,1,1,1,q,1,1,1,q) $ and three lower-diagonal elements $
{R\sp {21}}_{12} = {R\sp {31}}_{13} = {R\sp {32}}_{23} = \w , $
from which we get eight solutions for RE1, only two of them
have non-vanishing determinants
$$
K\sp {(1)}=\pmatrix{\a11& & \cr  &\a22& \cr  & &\a33\cr} , \quad
K\sp {(2)}=\pmatrix{\a11&\a12&\a13\cr   &f_{22}&f_{23}\cr
                     & &f_{33}\cr} , \quad  X=1+q\sp {-1} \eqn\sthre
$$
\noindent where the functions $f_{ij}$ depend on the parameters in the
first row $ f_{ii}=a_{1i}\sp 2/\a11 X\sp 2 $,\ $ f_{ij}=a_{1i} a_{1j}/\a11 X $.
By comparing with \spon\  we see that these solutions are rather different.
The other six solutions can obviously  be understood as an embedding of
the \sl2 solutions into \sl3 ones
$$  \eqalign{
    K\sp {(3)}&=\pmatrix{ &\a12& \cr  -q \a12& & \cr  & & \cr} ,  \cr
    K\sp {(4)}&=\pmatrix{ & &\a13\cr  & & \cr  -q \a13& & \cr} ,  \cr
    K\sp {(5)}&=\pmatrix{ & & \cr  & &\a23\cr   &-q \a23& \cr} ,  \cr}
       \qquad   \eqalign{
    K\sp {(6)}&=\pmatrix{\a11&\a12& \cr   &\a22& \cr   & & \cr} , \cr
    K\sp {(7)}&=\pmatrix{\a11& &\a13\cr   & & \cr   & &\a33\cr} , \cr
    K\sp {(8)}&=\pmatrix{ & & \cr   &\a22&\a23\cr   & &\a33\cr} . \cr}
\eqn\ksix   $$
It is interesting to note the difference between $K\sp{(2)}$ with
$a_{13}=0$ and $K\sp{(6)}$. In the latter $a_{22}$ is a free parameter
whereas in the former $f_{22}$ is not.

The embeddings of the lower dimensional solutions into the higher dimensional
ones are a general feature which we encounter again and again
for higher N.  For example, if we consider \sl4 with $R$-matrix given
by the following expression
$ R={\rm diag}(q,1,1,1,1,q,1,1,1,1,q,1,1,1,1,q) $ plus six
non-vanishing lower-diagonal elements
${R\sp {ij}}_{ji} = \w$, $1\le j<i\le4$,
then we find 13 solutions among which only three have non-vanishing
determinant.  Two of them are given by
$$
K\sp {(1)}=\pmatrix{\a11& & & \cr  &\a22& & \cr  & &\a33& \cr  & &
                         &\a44\cr} ,  \quad
K\sp {(2)}=\pmatrix{\a11&\a12&\a13&\a14\cr   &f_{22}&f_{23}&f_{24}\cr
            & &f_{33}&f_{34}\cr  & & &f_{44}\cr} , \eqn\sfour
$$
\noindent where the $ f_{ij} $ are the same functions as for \sl3.
The third solution is a double embedding of the metric type \sl2
solution into \sl4
$$
K\sp {(3)}=\pmatrix{ &\a12& & \cr  -q\a12& & & \cr  & & &\a34\cr
                       & &-q\a34& \cr} ,   \eqn\eq
$$
\noindent and all others are degenerate embeddings of \sl2. Of course,
there are a lot of truncations of $ K\sp {(1)} $ and $ K\sp {(2)} $ obtained
by putting to zero some of their elements, among them are also several
\sl3 embeddings.

\noindent Clearly, the solutions  $ K\sp {(1)} $, $ K\sp {(2)} $ and $K\sp{(3)}
$
suggest that
this structure is a general feature of \sl{N} for $ N>2 $.
However, \sl2 is special in that its solutions are less restricted.

To analyze \sl{N} we use the expression of the $R$-matrix in its
fundamental representation in terms of matrix units $ (e_{ij})_{ab} =
\delta_{ia}\delta_{jb} $
$$
R=q\sum_i e_{ii}\otimes e_{ii} + \sum_{i\not=j} e_{ii}\otimes
e_{jj} +\w\sum_{i>j} e_{ij}\otimes e_{ji} .  \eqn\RNfun
$$
\noindent We find that RE1 is
equivalent to the following set of equations for the  entries $ a_{ij} $ of
$K$ which are c-numbers :
$$ {\underline{i>j}}: \quad a_{ij}(a_{ij}+qa_{ji})=0 , \quad
                            a_{ii}a_{ij}=0 , \quad  a_{jj}a_{ij}=0, $$
$$ \eqalign{{\underline{{i>j>k}}: \quad }a_{jj}a_{ik}&=0 , \cr
    \phantom{\underline{{i>j>k}}: \quad }a_{ij}a_{ik}&=0 , \cr
    \phantom{\underline{{i>j>k}}: \quad }a_{ij}a_{ki}&=0 , \cr
    \phantom{\underline{{i>j>k}}: \quad }a_{ij}a_{jk}&=0 , \cr
    \phantom{\underline{{i>j>k}}: \quad }a_{ij}a_{kj}&=0 , \cr} \qquad
\eqalign{a_{ik}a_{jk}&=0 , \cr
         a_{ik}a_{kj}&=0 , \cr
         a_{ki}a_{jk}&=0 , \cr
         a_{ji}a_{ik}&=0 , \cr
         a_{ji}a_{jk}&=0 , \cr}  \qquad
\eqalign{Xa_{jj}a_{ki}-a_{ji}a_{kj}&=0 , \cr
         Xa_{ii}(a_{kj}-qa_{jk})-a_{ji}a_{ki}&=0 , \cr
         Xa_{kk}(a_{ji}-qa_{ij})-a_{ki}a_{kj}&=0 , \cr
           & \cr   X=1+q\sp {-1} ,& \cr}  $$
$$\eqalign{{\underline{{i>j>k>l}}: \quad }a_{ik}a_{jl}&=0 , \cr
    \phantom{\underline{{i>j>k>l}}: \quad }a_{il}a_{jk}&=0 , \cr} \qquad
\eqalign{a_{ik}a_{lj}&=0 , \cr
         a_{il}a_{kj}&=0 , \cr} \qquad
\eqalign{a_{ki}a_{jl}&=0 , \cr
         a_{li}a_{jk}&=0 , \cr} $$
\vskip-1cm
$$ \eqalign{\phantom{\underline{{i>j>k>l}}: \quad }a_{ij}a_{lk} -
                    a_{ji}a_{kl}&=0 , \cr
            \phantom{\underline{{i>j>k>l}}: \quad }a_{ij}a_{kl} + a_{li}a_{kj}
                  - a_{ji}a_{lk} + \w a_{ij}a_{lk}&=0 .  \cr} \eqn\set  $$
\noindent Let us examine some consequences of these equations for the
structure of possible solutions.

\noindent First, we note that there are no equations containing only $a_{ii}$,
hence it follows that a diagonal $K$-matrix with arbitrary elements
like $K\sp {(1)}$ is a solution.
An explanation of this fact based on the two-sided comodule property is
straightforward. From the explicit form of the $sl_q(N)$ $R$-matrix in the
fundamental representation \RNfun\ one can show that $R$ and $R\sp{t_1}$
commute,
$$
[R,R\sp{t_1}]=0 , \eqn\RRtone
$$
which means that the $N$ dimensional unit matrix is a solution
of RE1. By applying the symmetry transformation \JKJ\
(or the generalization of \simplCom)
to the unit matrix solution we get $K\sp{(1)}$ in \sthre\ and \sfour\ and in
general an arbitrary diagonal $N\times N$ matrix as a solution of RE1 for
$sl_q(N)$ case. In other words all the free parameters in the diagonal
solutions
(\eg\ $K\sp{(1)}$ in \sthre\ and \sfour) can be \lq\lq gauged away\rq\rq.

\noindent Second, for an upper-triangular $K$-matrix the
above system of equations
reduces drastically to
$$
\eqalign{\underline{{i>j>k}}: \quad Xa_{kk}a_{ji}-a_{ki}a_{kj}&=0 , \cr
    \phantom{\underline{{i>j>k}}: \quad }Xa_{jj}a_{ki}-a_{ji}a_{kj}&=0 , \cr
    \phantom{\underline{{i>j>k}}: \quad }Xa_{ii}a_{kj}-a_{ji}a_{ki}&=0 ,\cr
   \underline{{i>j>k>l}}: \quad a_{li}a_{kj}-a_{ji}a_{lk}&=0 . \cr}\eqn\redueq
$$
\noindent Choosing $ a_{1k}, k=1,\ldots,N $ as independent variables,
it is a simple exercise to verify that the following generalization of
$K\sp {(2)}$ to \sl{N}
$$
\bigl\{a_{jj}={{a_{1j}\sp 2}\over{a_{11} X\sp 2}} , \quad
a_{ji}={{a_{1j}a_{1i}}\over{a_{11}X}} , \quad i>j>1 \bigr\} \eqn\geso
$$
\noindent indeed satisfies these four equations. It is interesting to
point out that in contrast to the \sl2\ case, all the free parameters
($a_{11},\ldots,a_{1N}$) of the above upper-triangular solution for
$sl_q(N)$ $N>2$ can be made to unity (\lq\lq gauged away\rq\rq) by the
transformation \JKJ. For example, for $K\sp{(2)}$ in \sthre\ we have
$$
K\sp{(2)} \to U K\sp{(2)} U\sp{t_1}, \quad U={\rm diag}(1/\sqrt{a_{11}},
\sqrt{a_{11}}/a_{12}, \sqrt{a_{11}}/a_{13}) .
$$
Therefore the two types of generic solutions $K\sp{(1)}$ and $K\sp{(2)}$ of
RE1 for $sl_q(N)$ $N>2$ have essentially no free parameter.

\noindent Third, it is easy to see that there are no lower-triangular
solutions, as in this case the first equation in \set\  reduces to
$ a_{ij}\sp 2=0, i>j $,
hence all entries of a lower-triangular $K$-matrix vanish.

\noindent Finally, if we do not put any restrictions on $K$ we get
non-triangular \sl2 type embeddings, the structure of which is
indicated by the very first equation in \set.

\section{$sl_q(N|M)$ cases}
Next we consider $R$-matrices related directly to supersymmetric algebras,
which are, however,  solutions of the non-graded Yang-Baxter equations by the
well known transformation ($p(j)=0,1$ is the parity of the index $j$)
$$
R\sp{ij}{}_{kl}(\,{\rm YBE}) =
(-1)\sp{p(i)p(j)}\,R\sp{ij}{}_{kl}(\,\hbox{graded YBE}) .
$$
We also consider the non-graded reflection equations only.
The simplest example is \sl{1|1} connected with \refmark\rKuld
$$ R=\pmatrix{q& & & \cr  &1& & \cr  &\w&1& \cr  & & &-q\sp {-1}\cr},
\eqn\sup
$$
\noindent which enforces nilpotency of two quantum group generators.
In contrast to the $sl_q(2)$ case, the RE1 for $sl_q(1|1)$ has only one
solution
$$
K\sp {(1)}=\pmatrix{\a11&\a12\cr  &f\cr} , \eqn\supsol
$$
\noindent where $ f = -\a12\sp 2 / \a11\w $.
Namely, the counterpart of the $sl_q(2)$ solution $\ep_q$ is lacking.
For completeness we include the quadratic algebra
defined by RE1 and above $R$-matrix, given by
$$
\eqalign{\al\be &= \be\al +\w \al\ga  ,  \cr
         \al\ga &= q\sp 2 \ga\al         ,  \cr
         \al\de &= \de\al + q\w\ga\be ,  \cr
         \be\ga &= -q\sp 2 \ga\be        ,  \cr}  \qquad
\eqalign{\be\de &= \de\be + \w\ga\de  ,  \cr
         \ga\de &= q\sp {-2}\de\ga       ,  \cr
         \be\sp 2  &= -\w\al\de          ,  \cr
         \ga\sp 2  &= 0                  .  \cr}  \eqn\Alc
$$
This gives an interesting example of the deformation of two Grassmannian
quantities $\be$ and $\ga$, which have $\be\sp2=\ga\sp2=\be\ga+\ga\be=0$ in the
classical limit, \ie $q=1$. For $q\neq1$,  $\be\sp2$ and $\be\ga$
relations are
deformed whereas $\ga\sp2$ remains vanishing.
No linear central element $c_1$ exists for this algebra.
It is obvious that we get a trivial representation of the algebra by setting
$\ga=0$, which is given by the above solution $K\sp{(1)}$ \supsol.
It is clear that an equivalent formulation of \sl{1|1} can be obtained
from the above $R$-matrix by the transformation $ q=-1/q' $, which does
not affect $ \w $.

The three dimensional case, \ie\ $sl_q(N|M)$, $N+M=3$ is described by
the $R$-matrix $ R = {\rm diag}(\al,1,1,1,\be,1,1,1,\ga) $ plus three more
non-vanishing elements $ {R\sp {21}}_{12}={R\sp {31}}_{13}={R\sp {32}}_{23}=\w
$, and $\alpha,\beta,\gamma\in\{q,-q\sp {-1}\}$. There are eight different
solutions of the non-graded Yang-Baxter-equation for different choices of
$(\al,\be,\ga)$. However, due to the
transformation $ q=-1/q' $ there are only four independent ones.
Clearly, one of them is the \sl3 case discussed before (when $
\alpha=\beta=\gamma $), the others describe \sl{2|1}.
We do not discuss all of them as they are very similar, but as
an illustration choose $\alpha=\beta=-1/\gamma=q$. This gives three
independent solutions of RE1, an upper-triangular one
$$
K\sp {(1)}=\pmatrix{\a11&\a12&\a13\cr  &f_{22}&f_{23}\cr  & &f_{33}\cr} ,
\eqn\eq
$$
\noindent with $ f_{22}=\a12\sp 2/\a11X\sp 2, f_{23}=\a12\a13/\a11X,
f_{33}=-\a13\sp 2/\a11\w $, and two further solutions
$$
K\sp {(2)}=\pmatrix{ &\a12& \cr  -q\a12& & \cr  & &\a33\cr} , \quad
K\sp {(3)}=\pmatrix{\a11&\a12& \cr  &\a22& \cr  & & \cr} . \eqn\eq
$$
\noindent We obtain less solutions than for \sl3, but still they are
rather similar.
Especially, comparing $K\sp {(1)}$ of \sl{1|1} and \sl{2|1} with
their \sl2 and \sl3 counterparts we note that only the element in the
lower right corner is changed. If we describe \sl{N-1|1} case by the
\sl{N} $R$-matrix just replacing the element $ {R\sp {NN}}_{NN}=q $ by $
{R\sp {NN}}_{NN}=-1/q $, then it seems that there is a general upper-triangular
solution for \sl{N-1|1}, given
precisely by \geso\  except
that $ a_{NN}=a_{1N}\sp 2/\a11X\sp 2 $ is changed to $
a_{NN}=-a_{1N}\sp 2/\a11\w $. At least, we have checked this explicitly
for \sl{3|1}. Obviously $K\sp{(3)}$ above corresponds to $K\sp{(6)}$ in
\ksix\ of $sl_q(3)$. An interesting phenomenon is that $K\sp{(2)}$ above has
one more free parameter ($a_{33}$) than its \sl3\ counterpart, $K\sp{(3)}$ in
\ksix.

\section{Multiparameter $R$-matrices}
We now proceed to a different class of models, namely multiparameter
deformations. It will be interesting to see how previous results for
$K$ are affected by additional parameters. As already suggested by the
supersymmetric examples we will note that the more complicated the
$R$-matrix, the less structure the $K$-matrix usually can have.

\noindent We will discuss first $gl_{p,q}(2)$ described by the $R$-matrix
\REF\rDMMZ{\DMMZ} \REF\rSud{\Sud} \REF\rSWZ{\SWZ} \refmark{\rKuld-\rSWZ}
$$
R_{p,q}=\pmatrix{q& & & \cr  &p& & \cr  &\w&p\sp {-1}& \cr  & & &q\cr}, \eqn\eq
$$
\noindent which can be obtained from the 1-parameter $R$-matrix using
the general transformation \traa
$$
R_{p,q} = U_1\sp \p R_q U_2\sp {-1} , \qquad
U=\pmatrix{p\sp {1/2}& \cr  &p\sp {-1/2}\cr} . \eqn\twtra
$$
This $R$-matrix does not satisfy the conditions in \peR\ so we have to
use the generalized reflection equation \rethree\ here and in the
multiparameter case in general, still we refer to that reflection
equation also as RE1 since it is related to the same comodule
structure. The above transformation does not leave RE1 invariant, in
contrast to RE2, hence we expect different solutions compared to the
1-parameter case. Indeed we find four solutions, one of them is of
the quantum metric type as in \sol\
$$
K\sp {(1)}=\pmatrix{ &\a12\cr  -pq\a12& \cr} , \eqn\soltwa
$$
whereas the others all have vanishing determinant
$$
K\sp {(2)}=\pmatrix{\a11& \cr  & \cr} , \quad
K\sp {(3)}=\pmatrix{ &\a12\cr  & \cr} , \quad
K\sp {(4)}=\pmatrix{ & \cr  &\a22\cr} . \eqn\eq
$$
As in the 1-parameter case $K\sp {(1)}$ is stable under the comodule
transformation  up to the quantum determinant
$det_{p,q}T=ad-pqbc=da-p\sp {-1}q\sp {-1}cb$ which is not central for the
2-parameter quantum group $GL_{p,q}(2)$, i.e. we have $T K\sp {(1)} T\sp t =
(det_{p,q}T) K\sp {(1)}$.
In particular  the quadratic algebra defined by RE1 will not be
isomorphic to the 1-parameter algebra. It is given by
$$
\eqalign{\al\be &= p\sp {-2}\be\al + p\sp {-1}\w \al\ga ,  \cr
         \al\ga &= p\sp {-2}q\sp 2\ga\al            ,  \cr
         \al\de &= p\sp {-4}\de\al + p\sp {-2}\w(q\be\ga + p\sp {-1}\ga\sp 2) ,
\cr}
\qquad  \eqalign{\be\ga &= \ga\be   ,  \cr
                 \be\de &= p\sp {-2}\de\be + p\sp {-1}\w \ga\de ,  \cr
                 \ga\de &= p\sp {-2}q\sp 2\de\ga  ,  \cr}\eqn\Ald
$$
\noindent and for $p=1$ we get the 1-parameter algebra \Ala. For this
algebra there exists no central element  which is linear in the elements of
$K$.
However, $c_1(K)=\be-qp\sp{-1}\ga$ has simple relations: $c_1\al=p\sp2\al c_1$,
$c_1\de=p\sp{-2}\de c_1$, $[c_1,\be]=[c_1,\ga]=0$ and transforms homogeneously
$c_1(K_T)=(det_{p,q}T)c_1(K)$.

We now take a look at $ gl(3) $ which has  a 4-parameter deformation.
We find that the structure of constant solutions is completely
analogous to the $gl_{p,q}(2)$ case. The $R$-matrix is given by
$R={\rm diag}(q,p_1,p_2,p_1\sp {-1},q,p_3,p_2\sp {-1},p_3\sp {-1},q)$ plus the
non-vanishing elements $ {R\sp {21}}_{12} = {R\sp {31}}_{13} = {R\sp {32}}_{23}
= \w $ \refmark{\rDMMZ,\rSud}. It is worthwhile to point out that these
multiparameter constant solutions of the YBE
were already contained implicitly in \Ref\rPeS{\PeS} written some
ten years ago
(cf. formulae (25) and (26) therein). This was also  noted recently
in \Ref\rOY{\OY}. Unlike in the $N=2$ case
it cannot be obtained by such a simple transformation as \traa,
because in this way it is possible to add only $(N-1)$ independent
parameters to the 1-parameter $R$-matrix of \sl{N}. Anyway, in total
we have nine solutions, and three of them are generalized
$gl_{p,q}(2)$ embeddings
$$
K\sp {(1)}=\pmatrix{ &\a12& \cr  -p_1q\a12& & \cr  & & \cr} , \qquad
K\sp {(2)}=\pmatrix{ & &\a13\cr  & & \cr  -p_2q\a13& & \cr} ,
$$
$$
K\sp {(3)}=\pmatrix{ & & \cr  & &\a23\cr  &-p_3q\a23& \cr} .  \eqn\eq
$$
The other six solutions have one non-zero element only, located in the
upper-triangle, i.e. contained in the set
$\{a_{ij},i \leq j\}$. Looking at the $N=2,3$ solutions
one gets an idea what solutions for $N\geq4$ one might expect in
multiparameter cases.

\section{RE2}
Finally we look for constant solutions of the second reflection
equation \REtw\ or \reeqtw. We shall not go through the complete
analysis again,
but just pick up a few basic examples. We have already discussed the
\sl2 case in detail in section two. So let us proceed to \sl3 using
the $R$-matrix given at the beginning of section 3.2. RE2 is given by
$$
R K_1 \Rt K_2 = K_2 R K_1 \Rt ,
$$
and it has seven independent solutions for \sl3, two of them have
non-vanishing determinants
$$
K\sp {(1)}=\pmatrix{\a11& & \cr  &\a11& \cr  & &\a11\cr} , \qquad
K\sp {(2)}=\pmatrix{ & &\a13\cr  &\a22& \cr  g_{31}& &\a33\cr}  .  \eqn\ksolthr
$$
where the function $g_{31}$ is given by $g_{31}=\a22(\a22-\a33)/\a13$.
Again, as for RE1 we get a triangular 3-parameter solution and a
diagonal one, the latter is just the identity matrix which will always
be a solution, as is evident from  the structure of RE2.
However, as for RE1 we cannot transform away all two independent
parameters (disregarding an overall scaling parameter) by similarity
transformation \JKmJ\ but only one and the other one remains.

For \sl4 with $R$-matrix given below \ksix\ we get three solutions
with non-vanishing determinant. Besides the unity solution $K\sp{(1)}$
there are the following ones
$$
K\sp {(2)}=\pmatrix{ & & &\a14\cr  & &\a23& \cr  &\a32&\a33& \cr
                 g_{41}& & &\a33\cr} , \qquad
K\sp {(3)}=\pmatrix{ & & &\a14\cr  &\a22& & \cr  & &\a22& \cr
                 \a41& & &g_{44}\cr} ,  \eqn\ksolfour
$$
where the functions $ g_{14} $ and $ g_{44} $ are of quite different
forms, namely $ g_{41}=\a23\a32/\a14 $ and $
g_{44}=\a22-(\a14\a41)/\a22 $. They are also different from the
corresponding function in the \sl3 case above.
Besides them we found 14 independent solutions, many of
them also had three parameters as in \ksolfour.
The solutions $K\sp{(2)}$ in \ksolthr\ and \ksolfour\ strongly suggest the
generic solution of RE2 for the fundamental representation of $sl_q(N)$ having
the counter diagonal form,
$$
K \simeq \CD, \qquad (\CD)_{ij}= \delta_{i,\,N+1-j}. \eqn\couDi
$$
By using several properties of \CD\ ($e_{ij}$ is the matrix unit used in
\RNfun),
$$
\CD\sp2 = 1, \quad  e_{lm}\CD =\CD e_{\bar l \bar m}, \quad \bar m=N+1-m ,
$$
one can prove for the $R$-matrix in the fundamental representation
of $sl_q(N)$ the following commutation relation
$$
[ R, R_{\CD}] = 0,\qquad  R_{\CD} =\CD_2 R\CD_2 =\CD_1 \Rt \CD_1, \eqn\RhatR
$$
which corresponds to \RRtone\ in the RE1. It is easy to show that \CD
satisfies RE2 by using \RhatR. By applying the
transformation \JKmJ\ , we get a counter diagonal solution
with $N/2$ (or $(N-1)/2$) arbitrary parameters.

Next we treat again \sl{1|1} with $R$-matrix \sup, which gives the
following quadratic algebra for RE2
$$
\eqalign{\al\be &= q\sp {-2}\be\al                       ,  \cr
         \al\ga &= q\sp 2 \ga\al                         ,  \cr
         \al\de &= \de\al                             ,  \cr
         \be\ga &= -q\sp 2 \ga\be + q\w(\al\sp 2-\al\de)    ,  \cr}  \qquad
\eqalign{\be\de &= \de\be + q\w\al\be                 ,  \cr
         \ga\de &= \de\ga -q\w\ga\al                  ,  \cr
         \be\sp 2  &= 0                                  ,  \cr
         \ga\sp 2  &= 0                                  .  \cr}  \eqn\Ale
$$
The central element linear in the entries of $K$ is given by
$c_1=\al-\de$. It is obvious that this algebra is not isomorphic to
the algebra \Alc. This can be seen also  by the forms of
the constant solutions
$$
K\sp {(1)} \simeq \pmatrix{1& \cr  &1\cr}  ,  \qquad
K\sp {(2)}=\pmatrix{ & \cr   &\a22\cr}  ,  \eqn\eq
$$
where $K\sp {(1)}$ clearly cannot be transformed to the corresponding
solution $K\sp {(1)}$ in \supsol\ of RE1.

Finally some remarks on the multiparameter cases are in order.
For RE2 the two-sided comodule property requires
no conditions such as \peR\ on the $R$-matrices so that we can use
the same equation RE2 for multi-parameter cases, too. As stated before,
transformation \traa\ with an arbitrary
diagonal matrix $U$ leaves RE2 invariant for the $R$-matrices
related to \sl{N}.
Especially, the invariance under the transformation \twtra, which yields the
2-parameter $R$-matrix, means that the reflection equation for this
$R$-matrix reduces to the reflection equation for the 1-parameter
$R$-matrix. This is an important difference between RE1 and RE2.
A similar statement holds for $gl_{p_i,q}(3)$. Using transformation
\traa\ it is possible to obtain a 3-parameter deformation of a
special form which is completely equivalent to the 1-parameter case
\sl3. However, the general 4-parameter deformation
\refmark{\rDMMZ,\rSud}\ cannot be obtained
by a transformation \traa\  applied to the 1-parameter
$R$-matrix. It has three rather uninteresting solutions with vanishing
determinant plus the  unity solution. This proves that the
quadratic algebras defined by RE1 and RE2 for the $gl_{p_i,q}(3)$
$R$-matrix in fact are not isomorphic.

\chapter{Conclusions}

In this paper we have studied various constant solutions of the reflection
equations associated with a wide class of known $R$-matrices.
Many interesting properties of the reflection equations have been  uncovered
through the examination of these explicit solutions, some of which are
obtained by a formula manipulation program on a computer.
The study of constant solutions of the reflection equations has also
revealed rich algebraic structures which are directly related but not
identical with the quantum groups.
To name some of them, the quantum homogeneous spaces, generalization
of the braid groups, non-commutative differential geometry on quantum groups
\refmark{\rZuma,\rAF} and the fusion procedures for the $K$-matrices
\refmark{\rSklb,\rMN}, etc. In particular, the relations \Alb\ of the
quadratic algebra $\AA_2$ coincide with the relations of
the \lq\lq coordinates of the q-deformed Minkowski space\rq\rq
\Ref\rCWSSW{\CWSSW\nextline\SmWZ} and the central elements \centwo\
correspond to the time and the invariant length.

A variety of constant solutions presented here are expected to give
good starting points for constructing such objects and other possible
applications.
As another application of the constant solution $K$, let us mention the
construction of local fields in terms of exchange algebra fields.
A direct generalization into this direction with higher rank groups seems to
require the non-standard $R$-matrices associated with non-affine Toda
field theory
\Ref\rBDF{\CGa\nextline\BDF} for which the analysis turns out highly
non-trivial.

One could go on exploring various aspects of reflection equations on many
fronts. For example, understanding the reflection equations for the
$R$-matrices related to the quantum orthogonal $SO_q(N)$ and symplectic
$Sp_q(2N)$ groups
would be a nice challenge. These groups are characterized , in addition to
\rtt,
by $T C T\sp t = C $ with the matrix $C$ of the invariant quadratic forms
\refmark\rFRT.
These  constant matrices $C$ satisfy  the corresponding RE1.
The simplest example would be the matrix $\varepsilon_q$ in \sol\ for the
$sl_q(2)$ which may be considered as $sp_q(2)$.
Now that  a complete list of constant solutions to the YBE in two
dimensions (\ie\ dim $V_1$ = dim $V_2$ = 2)
\Ref\rHie{\Hie} is available,  one might be tempted to a systematic  study
of the
quadratic algebras and their one-dimensional representations (i.e.
constant solutions) of the reflection equations related with lower dimensional
quantum groups.
It would be nice to have connections between the constant solutions and
the spectral parameter dependent solutions of RE as in the cases of the YBE.

\vfill\eject

\refout
\end